\documentclass[a4paper,12pt]{article}
\usepackage{amssymb,amsmath,amsthm}
\usepackage{bm}%
\usepackage{ascmac}
\usepackage{mathrsfs}
\usepackage{stmaryrd}
\usepackage{comment}
\usepackage{xcolor}

\colorlet{mycol}{black}
\usepackage{hyperref}
\hypersetup{
	colorlinks=true,
	citecolor=blue,
	linkcolor=blue,
	urlcolor=blue,
}
\usepackage{arydshln}
\usepackage{braket}
\usepackage{tikz}
	\usetikzlibrary{arrows.meta,calc}
\usepackage{xcolor}

\def\nn{\notag}
\def\Z2{\mathbb{Z}_2^2}
\def\g{\mathfrak{g}}
\def\h{\mathfrak{h}}

\def\DP#1#2{\hat{#1}\cdot \hat{#2}}
\def\PH#1#2{(-1)^{\hat{#1}\cdot \hat{#2}}}
\def\alg{$\Z2$-$\osp(1|2)$}

\def\TL{\mathcal{L}}
\def\Ap{\mathscr{A}_+}
\def\Am{\mathscr{A}_-}
\def\Bp{\mathscr{B}_+}
\def\Bm{\mathscr{B}_-}
\def\osp{\mathfrak{osp}}
\def\sl{\mathfrak{sl}}

\numberwithin{equation}{section}

\title{
Integrable $\mathbb{Z}_2^2$-graded super-Liouville Equation and Induced $\mathbb{Z}_2^2$-graded super-Virasoro Algebra}
\author{Naruhiko Aizawa${}^{1}$\thanks{{E-mail: {\it aizawa@omu.ac.jp}} (corresponding author)}, \quad 
	Ichi Fujii${}^{1}$\thanks{E-mail: \textit{sq25482d@st.omu.ac.jp}}, \quad
	Ren Ito${}^{1}$\thanks{E-mail: \textit{sd22709y@st.omu.ac.jp}}, \\[5pt]
	Toshiya Tanaka${}^{1}$\thanks{E-mail: \textit{sd23429y@st.omu.ac.jp}} \quad and\quad
	Francesco Toppan${}^{2}$\thanks{{E-mail: {\it toppan@cbpf.br}}}
}
\date{\today}

\begin{document}
	
	\maketitle
	\thispagestyle{empty}
	\begin{center}
		{\small{	${}^{1}$\textit{Department of Physics, Graduate School of Science,
					\\
					Osaka Metropolitan University, Sugimoto Campus,
					\\
					Osaka 558-8585, Japan.}

				\bigskip
				${}^{2}$\textit{
					CBPF, Rua Dr. Xavier Sigaud 150, Urca,
					\\
					cep 22290-180, Rio de Janeiro (RJ), Brazil.
				}
		}}
	\end{center}

	\begin{abstract}
		We present a framework for enlarging the construction of $\Z2$-graded classical Toda theory from the class of $\Z2$-graded Lie algebras to the class of $\Z2$-graded Lie superalgebras. This scheme is applied to derive a $\Z2$-graded extension of the super-Liouville equation based on a $\Z2$-graded extension of $\osp(1|2).$  The mathematical tools employed in this work are a $\Z2$-graded version of the zero-curvature formalism and of the Polyakov's soldering procedure. It is demonstrated that both methods yield the same $\Z2$-graded super-Liouville equation. An algebraic construction of solutions to the resulting equations is also presented, together with their B\"acklund transformations. Furthermore, three distinct new $\Z2$-graded extensions of the super-Virasoro algebra are obtained via Hamiltonian reduction of the WZNW currents defined for $\Z2$-$\osp(1|2).$ 
	\end{abstract}
	
	\clearpage
	\setcounter{page}{1}
\section{Introduction}

A general framework for constructing integrable $\Z2$-graded extensions ($\Z2 := \mathbb{Z}_2 \times \mathbb{Z}_2$) of classical, two-dimensional Toda and conformal affine Toda models has been presented in \cite{aiktt}. A key feature of this construction is that it is based on a $\Z2$-graded  Lie algebra (also referred to as a $\Z2$-graded color Lie algebra) structure \cite{{rw1,rw2,Ree,sch}}; it implies that the resulting models contain fields obeying a special type of parabosonic statistics \cite{{GrJa},{yaji},{tol1},{kuto},{top2},{stvdjclass}}.
The construction was carried out for a $\Z2$-graded extension of $\sl(2), $ denoted by $\Z2$-$\sl(2), $ and its affine extension, resulting in integrable graded extensions of the Liouville and sinh-Gordon equations. 
Furthermore, by enlarging the Polyakov's soldering procedure \cite{polyak} to a $\Z2$-graded setting, an affine version of $\Z2$-$\sl(2) $ with a central extension was recovered as a Poisson-Lie algebra. 
It was also shown that a $\Z2$-extension of the Virasoro algebra naturally arises from a Hamiltonian reduction. \par
In this paper we extend the analysis of \cite{aiktt} to  a $\Z2$-graded Lie superalgebra framework. This corresponds to the more general case where not only parabosons, but also parafermionic fields are present
\cite{{tol1},StoVDJ, {top1},StoVDJ3,{bfrt},{StoVDJosp}}. This is the most general setting to investigate $\Z2$-graded 
paraparticles because it extends, see e.g. \cite{{vas},{tol},{brdu},{AAD}}, the notion of supersymmetry with all its physical and mathematical implications. \par
We stress the non-trivial significance of taking both parabosons and parafermions into account. 
The experimentalists learned how to simulate
paraoscillators \cite{parasim} and even engineer them in the laboratory  \cite{paraexp} by using trapped ions. On the theoretical side it has been shown, see \cite{{top1},{top2}} and also \cite{{bfrt},{top3}}, that certain results implied 
by $\Z2$-graded paraparticles  cannot be reproduced by ordinary Bose-Fermi statistics.  The connection of the $\Z2$-graded parastatistics with the traditional \cite{Green,GrMe} approach has been established in \cite{{tol1},{StoVDJ}}.\par
In this paper the Polyakov's soldering procedure and the Lax pair formalism are applied to $\Z2$-graded extensions of Lie superalgebras. 
To illustrate this scheme we examine a $\Z2$-graded version of $\osp(1|2)$ which allows to construct a $\Z2$-graded extension of the ordinary super-Liouville equation presented in \cite{ChaiKu,LeZSavLei}. \par
The $\Z2$-graded extension of a Lie (super)algebra is in general not unique.  Many inequivalent $\Z2$-graded versions of $\osp(1|2)$ have been discussed so far \cite{rw2,GrJa,StoVDJ,StoVDJ3,StoVDJosp,AAD,RY}. 
The one considered in this work is  the ten-dimensional  $\Z2$-graded Lie superalgebra introduced in \cite{AAD} as a $\Z2$-graded extension of the superconformal mechanics. \par
Concerning the Lax pair formulation, the $\Z2$ grading shares some feature with the ${\cal N}=2$ supersymmetry; instead of a single Lax pair as in the ${\cal N}=1$  models \cite{{LezSav1},{LezSav2},{LezSav3},{ToppanSL},{ToppanZhang}}, the equations are derived by two conjugate Lax pairs like the $N=2$ super-Toda models  \cite{ivto} (the ${\cal N}=2$ super-Liouville theory was originally  introduced in \cite{ivkr}).\par
A genuine new feature of the $\Z2$-graded superalgebra construction, with respect to the \cite{aiktt} $\Z2$-graded parabosonic {\textit{algebra}} case, is that the Hamiltonian reduction of the WZNW currents induces a nontrivial, color, 
$\Z2$-graded extension of the super-Virasoro algebra. This is a new $\Z2$-graded superVirasoro algebra which differs from the one presented in \cite{zhe}; its four component fields have respective gradings  $(00,11,10,01)$ and scaling dimensions $(2,2,\frac{3}{2},\frac32)$. The (anti)symmetry properties of its brackets imply a nontrivial $\Z2$-graded Lie superalgebra extension of ${\cal N}=1$ superVirasoro. In contrast, the Hamiltonian reduction from $\Z2$-graded $\sl(2)$ 
produces, see \cite{aiktt}, the $(00,11)$-graded subalgebra which is isomorphic to an ordinary (not colored) Lie algebra.\par
It should be mentioned that the $\Z2$-graded extensions of integrable systems discussed here and in \cite{aiktt} reveal the existence of a broader class of integrable systems than those known so far; it is indeed obvious that the present scheme will be easily extended to  more complex gradings such as $ \mathbb{Z}_2^n := \mathbb{Z}_2 \times \mathbb{Z}_2 \times \cdots \times \mathbb{Z}_2 $ ($n$ times). 
We remark that alternative approaches to $\Z2$-graded integrability, based on $\Z2$-graded supersymmetric Lagrangians in two-dimensional spacetime, can be found in the literature \cite{bruSG,NARITT2D}.

\bigskip

This paper is organized as follows: in \S \ref{SEC:alg} we recall the definition of $\Z2$-graded  Lie superalgebra and introduce a $\Z2$-graded extension of $\osp(1|2)$ which we denoted by $\g$ for simplicity. We also present representations of $\g$ which will be used in our analysis. 
In \S \ref{SEC:Equatoion} the $\Z2$-graded super-Liouville equation is derived via the Polyakov's soldering procedure. This is achieved by considering the simplest setting sufficient to produce a nontrivial $\Z2$-graded equation.  It will be shown in \S \ref{SEC:CurrentAlg} that the Hamiltonian reduction \cite{polyak} for the current associated with $\g$ , supplemented by an additional gauge fixing inspired by \cite{sab1,sab2}, gives rise to  a $\Z2$-graded extension of the $\mathcal{N}=1$ super-Virasoro algebra. The $\Z2$-grading admits, for the currents with nontrivial grading, three distinct (anti)periodicity conditions on a ${\bf S}^1$ circle, leading to three distinct versions of the derived $\Z2$-graded super-Virasoro algebra. 
We show in \S \ref{SEC:ZCF} that the $\Z2$-super-Liouville equation, derived by the soldering procedure, can be formulated in terms of Lax operators. This formulation enables us to construct explicit solutions in terms of chiral superfields. 
In \S \ref{SEC:Backlund} two B\"acklund transformations are presented: one relating the the $\Z2$-super-Liouville equation to the free equation, and the other providing the auto-B\"acklund transformation for the $\Z2$-super-Liouville equation. 
A summary of the present work, together with further remarks and possible directions for future research, is given in the Conclusion.

\section{$\Z2$-graded extension of $\osp(1|2)$} \label{SEC:alg}
\setcounter{equation}{0}

\subsection{Definitions}

 Let us recall the definition of the $\Z2$-graded Lie superalgebra \cite{rw1,rw2}. 
Let $\mathfrak{g}$ be a vector space and $\hat{a}\equiv [a_1a_2]$ an element of $\Z2$. 
Suppose that $\mathfrak{g}$ is a direct sum of graded components 
\begin{equation}
	\mathfrak{g} = \bigoplus_{\hat{a} \in \Z2}  g_{\hat{a}}= g_{[00]} \oplus g_{[10]} \oplus g_{[01]} \oplus g_{[11]}.   
\end{equation}
If $\mathfrak{g}$ admits a bilinear operation (the graded Lie bracket), denoted by $ \llbracket \cdot, \cdot \rrbracket $ and satisfying the identities 
\begin{align}
	& \llbracket A_{\hat{a}}, B_{\hat{b}}  \rrbracket \in g_{\hat{a}+\hat{b}},
	\\
	& \llbracket A_{\hat{a}}, B_{\hat{b}} \rrbracket = -(-1)^{\hat{a}\cdot\hat{b}} \llbracket  B_{\hat{b}}, A_{\hat{a}} \rrbracket,
	\\
	& (-1)^{\hat{a}\cdot\hat{c}} \llbracket A_{\hat{a}}, \llbracket B_{\hat{b}}, C_{\hat{c}} \rrbracket \rrbracket + \PH{b}{a} \llbracket B_{\hat{b}}, \llbracket C_{\hat{c}}, A_{\hat{a}} \rrbracket \rrbracket + \PH{c}{b}\llbracket C_{\hat{c}}, \llbracket A_{\hat{a}}, B_{\hat{b}} \rrbracket \rrbracket  = 0, \label{Jacobi}
\end{align}
where $ A_{\hat{a}}, B_{\hat{a}}, C_{\hat{a}} $ are homogeneous elements of $\mathfrak{g}_{\hat{a}}$ and 
\begin{equation}
	\hat{a} + \hat{b} = [a_1+b_1 ,a_2+b_2] \in \Z2, \qquad 
	\hat{a} \cdot \hat{b} = a_1 b_1 + a_2 b_2 \in \mathbb{Z}_2,
\end{equation}
then $\mathfrak{g}$ is referred to as a $\Z2$-graded Lie superalgebra.

It is clear from the definition that the graded Lie brackets are realized by commutators and anticommutators as follows 
\begin{equation}
	\llbracket A_{\hat{a}}, B_{\hat{b}} \rrbracket
	= 
	\begin{cases}
		[A_{\hat{a}}, B_{\hat{b}}], & \DP{a}{b} = 0,
		\\[10pt]
		\{ A_{\hat{a}}, B_{\hat{b}} \}, & \DP{a}{b} = 1.
	\end{cases}
\end{equation}
If $\llbracket A, B \rrbracket = 0, $ we say that $A$ and $ B$ are $\Z2$-\textit{graded commutative}. 
It is also observed from the definition that $ \g$ has $\mathbb{Z}_2$-grading, too:
\begin{equation}
	\g = \g_0 \oplus \g_1, \qquad \g_0 := \g_{[00]} \oplus \g_{[11]}, \ 
	\g_1 := \g_{[10]} \oplus \g_{[01]}.
\end{equation}

For a given Lie (super)algebra, one may consider its $\Z2$-graded extension which  is generally not unique.  Many inequivalent $\Z2$-graded extensions of the Lie superalgebra $\osp(1|2)$ have been discussed so far \cite{rw2,GrJa,StoVDJ,StoVDJ3,StoVDJosp,AAD,RY}. 
In the present work, we consider one of them which was the ten-dimensional  $\Z2$-graded Lie superalgebra introduced in \cite{AAD} as a $\Z2$-extension of the superconformal mechanics. 
Its basis, their $\Z2$-gradings  and scaling dimensions (eigenvalue of the grading operator $ G:= \frac{1}{2}K_0$) are summarized in the table below:
\begin{equation}
	\begin{array}{c|cccc}
		& [00] & [10] & [01] & [11] 
		\\[5pt] \hline
		+1 & K_+ &  &  & L_+ 
		\\[5pt]
		+\frac{1}{2} &  & P_+ & Q_+ & 
		\\[5pt] \hdashline 
		0  & K_0 & & & L_0 
		\\[5pt] \hdashline
		-\frac{1}{2} & & P_- & Q_- &  
		\\[5pt]
		-1 & K_- & &  &  L_-
	\end{array}
	\label{BasisGrading}
\end{equation}
We keep using $\g$ to denote this $\Z2$-graded extension of $\osp(1|2).$ 
The non-vanishing defining relations of $\g$ are given, in terms of (anti)commutators, as follows. 

\noindent
$-$ $\g_0$-$\g_0$ sector:
\begin{alignat}{4}
	[K_0, K_{\pm}] &= \pm 2 K_{\pm}, & \quad [K_+, K_-] &= K_0, &\quad [L_0, L_{\pm}] &= \pm 2 K_{\pm}, &\quad  [L_+, L_-] &= K_0,
	\nn \\[3pt]
	[K_0, L_{\pm}] &= \pm 2 L_{\pm},  & [K_{\pm}, L_{\mp}] &= \pm L_0, & [L_0, K_{\pm}] &= \pm 2 L_{\pm},
\end{alignat}

\medskip\noindent
$-$ $\g_0$-$\g_1$ sector:
\begin{alignat}{3}
	[K_0, P_{\pm}] &= \pm P_{\pm},  &\qquad  [K_{\pm}, P_{\mp}] &= -P_{\pm},
	\nn \\[3pt]
	[K_0, Q_{\pm}] &= \pm Q_{\pm},  &\qquad  [K_{\pm}, Q_{\mp}] &= -Q_{\pm},
	\nn \\[3pt]
	\{ P_{\pm}, L_0 \} &= \pm i Q_{\pm},  & \{ P_{\pm}, L_{\mp}\} &= -i Q_{\mp}, 
	\nn \\[3pt]
	\{ Q_{\pm}, L_0 \} &= \mp i P_{\pm},  & \{ Q_{\pm}, L_{\mp}\} &= i P_{\mp},
\end{alignat}

\medskip\noindent
$-$ $\g_1$-$\g_1$ sector:
\begin{alignat}{4}
	\{ P_{\pm}, P_{\pm} \} &= \pm 2 K_{\pm}, &\quad \{P_+, P_- \} &= K_0, &\quad [P_{\pm}, Q_{\pm}] &= \pm 2i L_{\pm}, &\quad [P_{\pm}, Q_{\mp}] &= i L_0,
	\nn \\[3pt]
	\{ Q_{\pm}, Q_{\pm}\} &= \pm 2 K_{\pm}, & \{Q_+, Q_-\} &=K_0. & 
\end{alignat}
It is easy to see that the algebra $\g$ admits  the triangular decomposition;
\begin{align}
	\g &= \g_+ \oplus \h \oplus \g_-,
	\\[3pt]
	\g_+ &= \{  K_+, \ P_+, \ Q_+, \ L_+ \},
	\nn \\[3pt]
	\h &= \{  K_0, \ L_0 \},
	\nn \\[3pt]
	\g_- &= \{  K_-, \ P_-, \ Q_-, \ L_- \} \nn
\end{align}
where $\h$ is the Cartan subalgebra of $\g$.

Structure and some representations of $\g$ are studied in \cite{NAKA,FaFaJ,NAJS}.  

%
\subsection{Representations} \label{SubSec:Rep}

In order to study integrable models associated with $\g$, the representation of $\g$ induced from the standard ($\mathbb{Z}_2$-graded) $\osp(1|2)$ is useful. 
We denote the basis of $\osp(1|2)$ by
\begin{equation}
	[0] \quad H, \ E_{\pm}, \qquad \qquad [1] \quad F_{\pm}
\end{equation} 
which subject to the non-vanishing relations
\begin{alignat}{3}
	[H, E_{\pm}] &= \pm 2 E_{\pm}, & \qquad [E_+, E_-] &= H, &\qquad [H, F_{\pm}] &= \pm F_{\pm},
	\nn \\[5pt]
	\{F_+, F_-\} &= H, & \{ F_{\pm}, F_{\pm} \} &= \pm 2 E_{\pm}, & [E_{\pm}, F_{\mp}] &= -F_{\pm}.
\end{alignat}
Let $\ket{v_0} $ be the lowest weight vector of the fundamental representation of $\osp(1|2)$, i.e.,
\begin{equation}
	F_- \ket{v_0} = 0, \qquad H \ket{v_0} = -\ket{v_0}, \qquad \braket{v_0|v_0} = 1. 
\end{equation} 
Define $\ket{v_n} := F_+^n \ket{v_0}$, then 
\begin{alignat}{3}
	H \ket{v_n} &= (n-1) \ket{v_n}, &\quad  F_+ \ket{v_n} &= \ket{v_{n+1}}, & \quad F_+ \ket{v_2} &= 0,
	\notag \\
	F_-\ket{v_1} &= -\ket{v_0}, & F_-\ket{v_2} &= \ket{v_1}. 
\end{alignat}

Now we introduce the $ 4 \times 4 $ complexified quaternionic matrices
\begin{align}
	M_0 &:= \mathbb{I}_2 \otimes \mathbb{I}_2, \quad M_1 := \mathbb{I}_2 \otimes \sigma_1, \quad M_2 := \sigma_1 \otimes \sigma_2, \quad M_3 := \sigma_1 \otimes \sigma_3
\end{align}
which are defined by the $2\times 2$ Identity and the Pauli matrices
\begin{equation}
	\mathbb{I}_2 = \begin{pmatrix} 
		1 & 0 \\ 0 & 1
	\end{pmatrix}, 
	\qquad
	\sigma_1 = 
	\begin{pmatrix}
		0 & 1 \\ 1 & 0
	\end{pmatrix},
	\qquad
	\sigma_2 = 
	\begin{pmatrix}
		0 & -i \\ i & 0
	\end{pmatrix},
	\qquad
	\sigma_3 = 
	\begin{pmatrix}
		1 & 0 \\ 0 & -1
	\end{pmatrix}.
\end{equation}
They satisfy for $i,j=1,2,3$ the relations (the totally antisymmetric structure constant $\epsilon_{ijk}$ is normalized so that $\epsilon_{123}=1$):
\begin{equation}
	M_i M_j = \delta_{ij} M_0+ i \epsilon_{ijk} M_k.
\end{equation}
We identify the $\Z2$-grading of the matrices as follows:
\begin{equation}
	[00]  \; : \; M_0; \qquad  [10] \; : \; M_1 ; \qquad [01] \; : \; M_2; \qquad [11] \; : \; M_3.
\end{equation}
Then, $\g$ is realized by the matrices $M_k$ and $ \osp(1|2)$ generators as follows (Cf. \cite{aizt,aiktt}):
\begin{alignat}{4}
	K_0 &= M_0 \otimes H, & \qquad K_{\pm} &= M_0 \otimes E_{\pm}, & \qquad P_{\pm} &= M_1 \otimes F_{\pm}, & \qquad Q_{\pm} &= M_2 \otimes F_{\pm},
	\nn \\
	L_0 &= M_3 \otimes H, & L_{\pm} &= M_3 \otimes E_{\pm}. 
	\label{MatrixReal}
\end{alignat}

This realization gives the six-dimensional lowest weight representation of $\g.$ 
There are two linearly independent lowest weight vectors on which \eqref{MatrixReal} act. 
They are given by $ \ket{00} = M_0 \otimes \ket{v_0} $ and $ \ket{11} = M_3 \otimes \ket{v_0} $ and have the grading indicated. Is is easy to verify that
\begin{align}
	P_- \ket{00} &= Q_- \ket{00} = P_- \ket{11} = Q_-\ket{11} = 0,
	\notag \\[3pt]
	K_0 \ket{00} &= -\ket{00}, \quad K_0\ket{11} = -\ket{11},
	\quad
	L_0 \ket{00} = -\ket{11}, \quad L_0\ket{11} = -\ket{00} \label{LWvec}
\end{align}
and
\begin{alignat}{2}
	P_+ \ket{00} &=-i Q_+ \ket{11}, & Q_+ \ket{00} &= i P_+ \ket{11}, 
	\nn \\
	P_+ Q_+ \ket{\hat{a}} &= -Q_+ P_+ \ket{\hat{a}}, & \qquad 
	P_+^n Q_+^m \ket{\hat{a}} &= 0, \quad \hat{a} = 00, 11 
\end{alignat}
where $ n+m  \geq 3. $ 
Therefore, the $\Z2$-graded representation space induced on $\ket{00}, \ket{11}$ is six-dimensional and its basis is taken to be
\begin{alignat}{2}
	&[00] &; \quad \ket{1} &:= K_+ \ket{00}, \qquad \ket{2} := \ket{00}, 
	\notag \\
	&[11] &; \quad \ket{3} &:= L_+ \ket{00}, \qquad \ket{4} := \ket{11}, 
	\notag \\ 
	&[10] &; \quad \ket{5} &:= P_+ \ket{00}, 
	\notag \\
	&[01] &; \quad \ket{6} &:= Q_+ \ket{00}. \label{6DrepBasis}
\end{alignat}
The action of $\g$ on this space is readily obtained. 
Setting $\vec{v} :=( \ket{1}, \ket{2}, \ket{3}, \ket{4}, \ket{5}, \ket{6}),$ we obtain the followings:
\begin{align}
	\vec{v} & \stackrel{K_0}{\longrightarrow} (\; \ket{1}, -\ket{2}, \ket{3},-\ket{4},0, 0 \;),
	\notag \\
	\vec{v} & \stackrel{K_+}{\longrightarrow} (\; 0, \ket{1}, 0, \ket{3}, 0, 0 \;),
	\notag \\
	\vec{v} & \stackrel{K_-}{\longrightarrow} (\; \ket{2}, 0, \ket{4}, 0, 0, 0 \;),
	\notag \\
	\vec{v} & \stackrel{L_0}{\longrightarrow} (\; \ket{3}, -\ket{4}, \ket{1}, -\ket{2}, 0, 0 \;),
	\notag \\
	\vec{v} & \stackrel{L_+}{\longrightarrow} (\; 0, \ket{3}, 0, \ket{1}, 0, 0 \;),
	\notag \\
	\vec{v} & \stackrel{L_-}{\longrightarrow} (\; \ket{4}, 0, \ket{2}, 0, 0, 0 \;),
	\notag \\
	\vec{v} & \stackrel{P_+}{\longrightarrow} (\; 0, \ket{5}, 0, -i\ket{6}, \ket{1}, i\ket{3} \;),
	\notag \\
	\vec{v} & \stackrel{P_-}{\longrightarrow} (\; \ket{5}, 0, -i\ket{6}, 0, -\ket{2}, -i\ket{4} \;),
	\notag \\
	\vec{v} & \stackrel{Q_+}{\longrightarrow} (\; 0, \ket{6}, 0, i\ket{5}, -i\ket{3}, \ket{1} \;),
	\notag \\
	\vec{v} & \stackrel{Q_-}{\longrightarrow} (\; \ket{6}, 0, i\ket{5}, 0, i\ket{4}, -\ket{2} \;). 
	\label{ActionGenerators}
\end{align}
It follows the following matrix representation of $\g$:  
\begin{alignat}{2}
	K_0 &= \mathrm{diag}(1,-1,1,-1,0,0), & \qquad 
	K_+ &= \begin{pmatrix}
		\sigma_+ & 0 & 0 \\ 0 & \sigma_+ & 0 \\ 0 & 0 & 0
	\end{pmatrix},
	\notag \\
	K_- &= \begin{pmatrix}
		\sigma_- & 0 & 0 \\ 0 & \sigma_- & 0 \\ 0 & 0 & 0
	\end{pmatrix},
	& L_0 &= \begin{pmatrix}
		0 & \sigma_3 & 0 \\ \sigma_3 & 0 & 0 \\ 0 & 0 & 0
	\end{pmatrix},
	\notag \\
	L_+ &= \begin{pmatrix}
		0 & \sigma_+ & 0 \\ \sigma_+ & 0 & 0 \\ 0 & 0 & 0 
	\end{pmatrix},
	& 
	L_- &= \begin{pmatrix}
		0 & \sigma_- & 0 \\ \sigma_- & 0 & 0 \\ 0 & 0 & 0
	\end{pmatrix},
	\notag \\
	P_+ &= \begin{pmatrix}
		0 & 0 & \sigma_{11} \\ 0 & 0 & i\sigma_+ \\ \sigma_+ & -i\sigma_{22} &{\color{mycol}0}
	\end{pmatrix},
	& 
	P_- &= \begin{pmatrix}
		0 & 0 & -\sigma_- \\ 0 & 0 & -i\sigma_{22} \\ \sigma_{11} & -i\sigma_- & 0
	\end{pmatrix},
	\notag \\
	Q_+ &= \begin{pmatrix}
		0 & 0 & \sigma_+ \\ 0 & 0 & -i\sigma_{11} \\ \sigma_{22} & i \sigma_+ & 0
	\end{pmatrix},
	&
	Q_- &= \begin{pmatrix}
		0 & 0 & -\sigma_{22} \\ 0 & 0 & i\sigma_- \\ \sigma_- & i\sigma_{11} & 0
	\end{pmatrix}
\end{alignat}
where 
\begin{alignat}{2}
	\sigma_3 &= \begin{pmatrix}
		1 & 0 \\ 0 & -1
	\end{pmatrix},
	&\qquad 
	\sigma_+ &= \begin{pmatrix}
		0 & 1 \\ 0 & 0
	\end{pmatrix},
	\qquad 
	\sigma_- = \begin{pmatrix}
		0 & 0 \\ 1 & 0
	\end{pmatrix},
	\notag \\
	\sigma_{11} &= \begin{pmatrix}
		1 & 0 \\ 0 & 0
	\end{pmatrix},
	& \sigma_{22} &= \begin{pmatrix}
		0 & 0 \\ 0 & 1
	\end{pmatrix}.
\end{alignat}

\section{$\Z2$-super-Liouville equation by Polyakov's soldering} \label{SEC:Equatoion}
\setcounter{equation}{0}

We mimick the standard procedure of soldering for deriving \cite{polyak} the $\Z2$-graded version of the super-Liouville equation. We introduce the $\Z2$-graded  Lie group $\Z2$-$Osp(1|2)$ generated by the algebra \alg defined in \S\ref{SEC:alg}. A group element of $\Z2$-$Osp(1|2)$ is parametrized by 
\begin{align}
	g &= \exp(\alpha_{00} K_+ + \alpha_{11} L_+) \exp(\lambda_{10} P_+ + \lambda_{01} Q_+) \exp(\beta_{00} K_0 + \beta_{11} L_0) 
	\notag \\
	& \times \exp(\mu_{10}P_-  + \mu_{01}Q_-) \exp(\gamma_{00}K_- + \gamma_{11} L_-) 
	\label{groupelem}
\end{align}
where the group parameters $ \alpha, \beta, \lambda, \mu $ and $ \gamma $ are also $\Z2$-commutative and their grading is indicated by the suffix. 
In particular, $ \lambda_{10}, \lambda_{01}, \mu_{10} $ and $ \mu_{01}$ are nilpotent. 
Throughout this article, the suffices 00, 10, 01, 11 indicate the $\Z2$-grading of the associated quantities.  
We assume that the group parameters are superfields which are functions of the [00] and [10]-graded $\Z2$-commutative variables
\begin{equation}
	[00] \ x, \ \bar{x}, \qquad [10] \ \theta, \ \bar{\theta}. \label{10coordinates}
\end{equation}
Alternatively, one could assume that the parameters are superfields on [00] and [01]-graded variables. 
It is obvious that both assumptions lead to the same equation, so we consider only the case of [00] and [10]-graded variables. 

We introduce the [10]-graded covariant derivatives
\begin{equation}
	D = \partial_{\theta} + i \theta \partial_{x}, \qquad \bar{D} = \partial_{\bar{\theta}} + i \bar{\theta} \partial_{\bar{x}}
\end{equation}
which satisfy
\begin{align}
	\{ D, D \} = 2i  \partial_{x}, \qquad \{ \bar{D}, \bar{D} \} = 2i  \partial_{\bar{x}}, \qquad \{ D, \bar{D} \}  = 0. 
\end{align}
We introduce also the holomorphic and antiholomorphic WZNW-currents which are defined in terms of the group element \eqref{groupelem}:
\begin{equation}
	J(x,\theta) := D g\cdot g^{-1}, \qquad \bar{J}(\bar{x}, \bar{\theta}) := g^{-1} \bar{D} g. 
	\label{DEF:current} 
\end{equation}
By definition, the currents $J(x,\theta), \bar{J}(\bar{x}, \bar{\theta})$ are [10]-graded and take values in $\Z2$-$\osp(1|2).$  
Employing the matrix presentation \eqref{MatrixReal}, one may rearrange the components of the currents in terms of the  $\osp(1|2)$ generators.  
First, the group element \eqref{groupelem} is given by
\begin{equation}
	g = \exp( a \otimes E_+) \exp(b \otimes F_+) \exp(c\otimes  H) \exp(d\otimes F_-) \exp(f \otimes E_-)
\end{equation}
where the matrix valued fields $ a, b, c, d $ and $ f $ are defined by
\begin{alignat}{2}
	a &:= \alpha_{00} M_0 + \alpha_{11} M_3,
	& \qquad 
	b &:= \lambda_{10} M_1 + \lambda_{01} M_2,
	\notag \\
	c &:= \beta_{00} M_0 + \beta_{11} M_3, & 
	d &:= \mu_{10} M_1 + \mu_{01} M_2,
	\notag \\
	f &:= \gamma_{00} M_0 + \gamma_{11} M_3 \label{DEF:abc}
\end{alignat}
and due to the nilpotency of $ \lambda_{10}, \lambda_{01}, \mu_{10}$ and $ \mu_{01} $ we have the relations
\begin{equation}
	b^2 = d^2 = \{ b, d \} = 0. 
\end{equation}
It follows immediately that the currents \eqref{DEF:current} takes the following form:
\begin{align}
	J &= J_{++} \otimes E_+ + J_+ \otimes F_+ + J_0 \otimes H + J_- \otimes F_- + J_{--} \otimes E_-,
	\notag \\
	\bar{J} &= \bar{J}_{++} \otimes E_+ + \bar{J}_+ \otimes F_+ + \bar{J}_0 \otimes H + \bar{J}_- \otimes F_- + \bar{J}_{--} \otimes E_- 
	\label{CurrentOspbasis}
\end{align}
where
\begin{align}
	J_{++} &= -2e^{-2c}\big( Df - (Dd)d \big)a^2 -2e^{-c}(Dd)ab -2(Dc)a + (Db)b + Da,
	\notag \\[3pt]
	J_+ &= -e^{-2c}\big( Df - (Dd)d \big) ab -e^{-c}(Dd)a - (Dc)b + Db,
	\notag \\[3pt]
	J_0 &= e^{-2c} \big( Df - (Dd)d \big) a + e^{-c}(Dd)b + Dc,
	\notag \\[3pt]
	J_- &= e^{-2c}\big(Df- (Dd)d \big) b +e^{-c} Dd,
	\notag \\[3pt]
	J_{--} &= \big(Df - (Dd)d \big)e^{-2c}
\end{align}
and
\begin{align}
	\bar{J}_{++} &= e^{ -2c} \big( \bar{D}a + b \bar{D}b\big),
	\notag \\[3pt]
	\bar{J}_+ &= -e^{ -2c} \big( \bar{D}a + b \bar{D}b\big) d + e^{-c} \bar{D}b,
	\notag \\[3pt]
	\bar{J}_0 &= -e^{ -2c} \big( \bar{D}a + b \bar{D}b\big) f - e^{-c}(\bar{D}b)d + \bar{D}c,
	\notag \\[3pt]
	\bar{J}_- &= -e^{ -2c} \big( \bar{D}a  + b \bar{D}b\big)fd + e^{-c}(\bar{D}b) f - (\bar{D}c)d + \bar{D}d,
	\notag \\[3pt]
	\bar{J}_{--} &= -e^{ -2c} \big( \bar{D}a + b \bar{D}b\big)f^2 -2 e^{-c}(\bar{D}b) fd - 2(\bar{D}c)f  + (\bar{D}d)d + \bar{D}f.
\end{align}

According to \cite{polyak} we impose Hamiltonian constraints on the currents. 
Taking into account the grading and the matrix nature of the components of $ J, \bar{J}$, the appropriate constraints will be
\begin{alignat}{2}
	J_{--} &= J_0 = 0, & \qquad J_- &= M_1,
	\label{RductionConstraints} \\
	\bar{J}_{++} &= \bar{J}_0 = 0, & \bar{J}_+ &= M_1. \label{RductionConstraints2}
\end{alignat}
The constraints on $ J_{--}, J_{-} $ and $ J_0 $ give
\begin{equation}
	D c = -bM_1 \label{Sol1}
\end{equation}
and the constraints on $ \bar{J}_{++}, \bar{J}_+ $ gives
\begin{equation}
	\bar{D}b = e^c M_1. \label{Sol2}
\end{equation}
Acting $\bar{D}$ on \eqref{Sol1} and using \eqref{Sol2}, we obtain
\begin{equation}
	D \bar{D} c = e^c.
\end{equation}
Recalling the definition of $c$ in \eqref{DEF:abc}, it follows that 
\begin{equation}
	e^c = e^{\beta_{00} M_0 } e^{\beta_{11} M_3} = e^{\beta_{11}} (\cosh \beta_{11} M_0 + \sinh \beta_{11} M_3).
\end{equation}
Thus we obtain the following system of equations:
\begin{equation}
	D \bar{D} \beta_{00} = e^{\beta_{00}} \cosh \beta_{11}, 
	\qquad
	D \bar{D} \beta_{11} = e^{\beta_{00}} \sinh \beta_{11}.\label{SLsol}
\end{equation}

Due to the nilpotency of $\theta, \bar{\theta} $ one may expand the superfields into their components:
\begin{align}
	\beta_{00} &= \varphi_{00}(x,\bar{x}) + \theta \psi_{10}(x,\bar{x}) + \bar{\theta} \bar{\psi}_{10}(x,\bar{x}) + \theta \bar{\theta} F_{00}(x,\bar{x}),
	\notag \\
	\beta_{11} &=  \varphi_{11}(x,\bar{x}) + \theta \psi_{01}(x,\bar{x}) + \bar{\theta} \bar{\psi}_{01}(x,\bar{x}) + \theta \bar{\theta} F_{11}(x,\bar{x}).
	\label{PhiComponents}
\end{align}
Although we started with only $[10]$-graded superspace, the superfield $\beta_{11}$ ensures the existence of $[01]$-graded component fields so that the $\Z2$-super-Liouville equations \eqref{SLsol} are a system of coupled equations for $[00], [11], [10]$ and $[01]$-graded fields defined on two-dimensional spacetime. 
More explicitly, from the first equation in \eqref{SLsol} we get
\begin{align}
	\partial \bar{\partial} \varphi_{00} &= e^{\varphi_{00}} 
	\big( 
	\cosh \varphi_{11} (\psi_{10} \bar{\psi}_{10} - \psi_{01} \bar{\psi}_{01}- F_{00} )
	+
	\sinh \varphi_{11} (\psi_{10}\bar{\psi}_{01} - \psi_{01} \bar{\psi}_{10}  - F_{11})
	\big),
	\notag \\[3pt]
	i \bar{\partial} \psi_{10} &= e^{\varphi_{00}} (-\cosh \varphi_{11}\cdot \bar{\psi}_{10} + \sinh \varphi_{11} \cdot \bar{\psi}_{01}),
	\notag \\[3pt]
	i \partial \bar{\psi}_{10} &= e^{\varphi_{00}} (\cosh \varphi_{11}\cdot \psi_{10} - \sinh \varphi_{11} \cdot \psi_{01}),
	\notag \\[3pt]
	F_{00} &= -e^{\varphi_{00}} \cosh \varphi_{11},
	\label{CompEq1}
\end{align}
where $ \partial = \partial_{x}, \bar{\partial} = \partial_{\bar{x}} $  and from the second we get
\begin{align}
	\partial \bar{\partial} \varphi_{11} &= e^{\varphi_{00}} 
	\big( 
	\cosh \varphi_{11} (\psi_{10} \bar{\psi}_{01} - \psi_{01} \bar{\psi}_{10}- F_{11} )
	+
	\sinh \varphi_{11} (\psi_{10}\bar{\psi}_{10} - \psi_{01} \bar{\psi}_{01} - F_{00})
	\big),
	\notag \\[3pt]
	i \bar{\partial} \psi_{01} &= e^{\varphi_{00}} (-\cosh \varphi_{11}\cdot \bar{\psi}_{01} + \sinh \varphi_{11} \cdot \bar{\psi}_{10}),
	\notag \\[3pt]
	i \partial \bar{\psi}_{01} &= e^{\varphi_{00}} (\cosh \varphi_{11}\cdot \psi_{01} - \sinh \varphi_{11} \cdot \psi_{10}),
	\notag \\[3pt]
	F_{11} &= -e^{\varphi_{00}} \sinh \varphi_{11}. 
	\label{CompEq2}
\end{align}
It is observed that $ F_{00} $ and $ F_{11}$ are non-propagating auxiliary fields. 
Elimination of them gives the following equations of motion of $\varphi_{00}$ and $\varphi_{11}:$  
\begin{align}
	\partial \bar{\partial} \varphi_{00} &= e^{2\varphi_{00}} \cosh 2\varphi_{11}
	\notag \\
	&
	+e^{\varphi_{00}} 
	\big(  
	\cosh \varphi_{11} (\psi_{10} \bar{\psi}_{10} - \psi_{01} \bar{\psi}_{01} )
	+
	\sinh \varphi_{11} (\psi_{10}\bar{\psi}_{01} - \psi_{01} \bar{\psi}_{10} )
	\big),
	\notag \\[3pt]	
	\partial \bar{\partial} \varphi_{11} &= 
	e^{2\varphi_{00}} \sinh 2\varphi_{11}
	\notag \\
	&
	+
	e^{\varphi_{00}} 
	\big( 
	\cosh \varphi_{11} (\psi_{10} \bar{\psi}_{01} - \psi_{01} \bar{\psi}_{10} )
	+
	\sinh \varphi_{11} (\psi_{10}\bar{\psi}_{10} - \psi_{01} \bar{\psi}_{01})
	\big). 
	\label{ComponentEq2}
\end{align}

Setting $\varphi_{11} $ and $ \psi_{01}, \bar{\psi}_{01}$ (or $ \psi_{10}, \bar{\psi}_{10}$) equal to zero, the super-Liouville equations discussed in \cite{ChaiKu} are recovered. 
Setting all the fields with non-trivial grading zero, the Liouville equation is recovered. 
Therefore, the $\Z2$-super-Liouville equations constructed here are a natural generalization of the (super-)Liouville equation. 
The integrability of the system of equations \eqref{SLsol} is ensured by its formulation as a zero-curvature representation which we discussed in \S \ref{SEC:ZCF}. 
Following the method of Leznov and Saveliev, we also construct solutions to the equations presented in \S \ref{SEC:LSA}.

\section{$\Z2$-graded super-Virasoro algebras} \label{SEC:CurrentAlg}
\setcounter{equation}{0}

In this Section we consider the current algebra associated with the currents given in \eqref{DEF:current}.  
Polyakov showed that the Virasoro algebra emerges from constraining the current algebra associated with $SL(2,\mathbb{R})$  \cite{polyak}. This construction was extended to $ Osp(1|2)$ by Sabra \cite{sab1,sab2}, where  the $ \mathcal{N} = 1 $ super-Virasoro algebra is obtained. In our case, a $\Z2$-graded generalization of the super-Virasoro algebra is found. This analysis is carried out within the framework of classical mechanics, i.e., making use of the Poisson brackets. 
Nevertheless, as in the case of $\Z2$-graded Liouville equation based on $\Z2$-$\sl(2)$ discussed in \cite{aiktt}, we observe the existence of a central term in the Poisson Lie algebra (an example, see \cite{Anomaly}, of a classical anomaly). 
 
Only the holomorphic sector will be considered here, as the treatment of the antiholomorphic sector proceeds in a similar fashion. 
The transformations of the component currents defined in \eqref{CurrentOspbasis} are induced from the left action of the group element: 
\begin{align}
	g &\to g' = g_{\epsilon} g,
	\notag \\
	g_{\epsilon} &= e^{\epsilon_{++}\otimes E_+}e^{\epsilon_+\otimes F_+}e^{\epsilon_0 \otimes H_+}e^{\epsilon_-\otimes F_-}e^{\epsilon_{--} \otimes E_-}
\end{align}
where $ \epsilon$'s are holomorphic functions and $ \epsilon_{\pm} $ are nilpotent and anticommute. 
Considering the infinitesimal transformation of $g$, one may obtain
\begin{align}
	\delta J_{++}
	& = -2\epsilon_{++}J_0 +2\epsilon_+J_+ +2\epsilon_0J_{++} +D\epsilon_{++},
	\notag\\[2pt]
	\delta J_+
	& = -\epsilon_{++}J_- +\epsilon_+J_0 +\epsilon_0J_+ -\epsilon_-J_{++} +D\epsilon_+,
	\notag\\[2pt]
	\delta J_0
	& = \epsilon_{++}J_{--} +\epsilon_+J_- +\epsilon_-J_+ -\epsilon_{--}J_{++} +D\epsilon_0,
	\notag\\[2pt]
	\delta J_-
	& = -\epsilon_+J_{--} -\epsilon_0J_- -\epsilon_-J_0 -\epsilon_{--}J_+ +D\epsilon_-,
	\notag\\[2pt]
	\delta J_{--}
	& = -2\epsilon_0J_{--} -2\epsilon_-J_- +2\epsilon_{--}J_0 +D\epsilon_{--}.\label{deltaJ}
\end{align}

Besides the constraints given in \eqref{RductionConstraints}, following \cite{sab1,sab2} we impose an additional gauge fixing 
\begin{equation}
	J_+ = 0. \label{gaugefixing}
\end{equation} 
Then, the infinitesimal transformations preserving \eqref{RductionConstraints} and \eqref{gaugefixing} reduce to a single parameter:
\begin{align}
	\epsilon_-
	& = \frac{1}{2}D\epsilon_{--}M_1,
	\qquad 
	\epsilon_0
	 =\frac{i}{2}\epsilon_{--}',
	\notag \\
	\epsilon_+
	& = \epsilon_{--}J_{++}M_1	-\frac{i}{2}D\epsilon_{--}'M_1,
	\notag \\
	\epsilon_{++} &= \frac{1}{2}(D\epsilon_{--}) J_{++} + \frac{1}{2}\epsilon_{--} DJ_{++} + \frac{1}{2}\epsilon_{--}''.
\end{align}
So that
\begin{equation}
	\delta J_{++} = \frac{3i}{2} \epsilon_{--}' J_{++} + (D\epsilon_{--}) DJ_{++} + i \epsilon_{--} J_{++}' + \frac{1}{2}D\epsilon_{--}'' \label{deltaJpp}
\end{equation}
where the prime stands for the derivative with respect to $x:$ $a' = \partial_{x}a. $ 
Recalling that $ J_{++} $ consists of $\Z2$-graded superfields and matrices $M_i,$ one may recover the $\Z2$-graded currents $J_{10} $ and $ J_{01}$ by
\begin{equation}
	J_{++}= J_{10}M_0+J_{01}M_3.
\end{equation}
$\epsilon_{--}$ is also parametrized as:
\begin{equation}
	\epsilon_{--} = \epsilon_{00} M_0 + \epsilon_{11} M_3.
\end{equation}
The transformations of $ J_{10} $ and $ J_{01}$ can be readily obtained from \eqref{deltaJpp}:
\begin{align}
	\delta J_{10}
	& = \frac{3i}{2}\left(\epsilon_{00}^{\prime}J_{10}-\epsilon_{11}^{\prime}J_{01}\right)
	+\frac{1}{2}\left[\left(D\epsilon_{00}\right)DJ_{10}
	+\left(D\epsilon_{11}\right)DJ_{01}\right]
	\nn
	\\
	& +i\epsilon_{00}J_{10}' -i\epsilon_{11}J_{01}'
	+\frac{1}{2}D\epsilon_{00}^{\prime\prime},\label{tr3:1}
	\\[5mm]
	\delta J_{01}
	& = \frac{3i}{2}\left(\epsilon_{00}^{\prime}J_{01}- \epsilon_{11}^{\prime}J_{10}\right)
	+\frac{1}{2}\left[\left(D\epsilon_{00}\right)DJ_{01}
	+\left(D\epsilon_{11}\right)DJ_{10}\right]
	\nn
	\\
	& +i\epsilon_{00}J_{01}' -i\epsilon_{11}J_{10}'
	+\frac{1}{2}D\epsilon_{11}^{\prime\prime}.\label{tr3:2}
\end{align}
By assigning the scaling dimension $ [x]=-1,$ the scaling dimensions of the remaining quantities are determined as follows:
\begin{equation}
	[\theta]=-\frac{1}{2},\qquad [J_{10}] = [J_{01}] = \frac{3}{2}, \qquad [\epsilon_{00}] = [\epsilon_{11}] = -1.
\end{equation}

 One may expand the currents and the parameters in their components:
\begin{alignat}{2}
	J_{10}(x,\theta)
	& =u_{10}(x)+\theta u_{00}(x),
	& \qquad 
	J_{01}(x,\theta)
	& =u_{01}(x)+\theta u_{11}(x),
	\nn \\
	\epsilon_{00}(x,\theta)
	& =\varepsilon_{00}(x)+\theta\varepsilon_{10}(x),
	&
	\epsilon_{11}(x,\theta)
	& =\varepsilon_{11}(x)+\theta\varepsilon_{01}(x).
\end{alignat} 
The transformations of the components are readily obtained
\begin{align}
	\delta u_{00} &= i\left(2 \varepsilon_{00}' u_{00} + \varepsilon_{00} u_{00}' + \frac{1}{2} \varepsilon_{00}''' \right) 
	+ i\big( 2\varepsilon_{11}' u_{11} + \varepsilon_{11} u_{11}' \big)
	\notag \\
	&+i \left(\frac{3}{2} \varepsilon_{10}' u_{10} + \frac{1}{2}\varepsilon_{10} u_{10}' \right)
	-i \left(\frac{3}{2} \varepsilon_{01}' u_{01} + \frac{1}{2}\varepsilon_{01} u_{01}' \right),
	\notag \\
	\delta u_{11} &= i\left(2 \varepsilon_{11}' u_{00} + \varepsilon_{11} u_{00}' + \frac{1}{2} \varepsilon_{11}''' \right) 
	+ i\big( 2\varepsilon_{00}' u_{11} + \varepsilon_{00} u_{11}' \big)
	\notag \\
	&+i \left(\frac{3}{2} \varepsilon_{10}' u_{01} + \frac{1}{2}\varepsilon_{10} u_{01}' \right)
	-i \left(\frac{3}{2} \varepsilon_{01}' u_{10} + \frac{1}{2}\varepsilon_{01} u_{10}' \right)	
	\label{Utransf1}
\end{align}
and
\begin{align}
	\delta u_{10} &= i\left( \frac{3}{2} \varepsilon_{00}' u_{10} + \varepsilon_{00} u_{10}'\right)
	-i\left( \frac{3}{2} \varepsilon_{11}' u_{01} + \varepsilon_{11} u_{01}'\right) 
	+ \frac{1}{2} \left( \varepsilon_{10} u_{00} + \varepsilon_{01} u_{11} + \varepsilon_{10}'' \right),
	\notag \\
	\delta u_{01} &= i\left( \frac{3}{2} \varepsilon_{00}' u_{01} + \varepsilon_{00} u_{01}'\right)
	-i\left( \frac{3}{2} \varepsilon_{11}' u_{10} + \varepsilon_{11} u_{10}'\right) 
	+ \frac{1}{2} \left( \varepsilon_{01} u_{00} + \varepsilon_{10} u_{11} + \varepsilon_{01}'' \right). \label{Utransf2}
\end{align}
It can be readily verified that the scaling dimension is determined as follows:
\begin{equation}
	[u_{00}]=[u_{11}]=2, \quad [u_{10}]=[u_{01}]=\frac{3}{2},\quad [\varepsilon_{00}]
	=[\varepsilon_{11}]=-1,\quad[\varepsilon_{10}]=[\varepsilon_{01}]=-\frac{1}{2}.
\end{equation}
It follows from these results that the currents $ u_{00}, u_{11}, u_{10} $ and $ u_{01} $ can be identified with a $\Z2$-graded generalization of the  Virasoro algebra. 
To see this explicitly, we need to introduce a correct Poisson brackets structure.

The infinitesimal transformations \eqref{Utransf1} and \eqref{Utransf2} may be reproduced by the Poisson bracket:
\begin{align}
	\delta u(x) &= \frac{1}{2\pi}\oint  dy \{K(y) , u(x)  \},
	\notag \\
	K(y)&:=k_1 \varepsilon_{00}u_{00} +k_2 \varepsilon_{11}u_{11} +k_3 \varepsilon_{10}u_{10} +k_4 \varepsilon_{01}u_{01},
	\label{TransbyGen}
\end{align} 
where $u(x)$ stands for the $\Z2$-graded component currents and the $k_i$ constants have to be determined. 
The $\Z2$-graded Poisson bracket is defined by the relations
\begin{align}
	\left\{u_{\hat{a}},\, u_{\hat{b}}\right\}
	& =-(-1)^{\hat{a}\cdot\hat{b}}\left\{u_{\hat{b}},\, u_{\hat{a}}\right\},\label{PB_def}
	\notag \\
	\left\{u_{\hat{a}}u_{\hat{b}},\, u_{\hat{c}}\right\}
	& =u_{\hat{a}}\left\{u_{\hat{b}},\, u_{\hat{c}}\right\}
	+(-1)^{\hat{b}\cdot\hat{c}}\left\{u_{\hat{a}},\, u_{\hat{c}}\right\}u_{\hat{b}},
	\notag \\
	\left\{u_{\hat{a}},\, \left\{u_{\hat{b}},\, u_{\hat{c}}\right\}\right\}
	& =\left\{\left\{u_{\hat{a}},\, u_{\hat{b}}\right\},\, u_{\hat{c}}\right\}
	+(-1)^{\hat{a}\cdot\hat{b}}\left\{u_{\hat{b}},\, \left\{u_{\hat{a}},\, u_{\hat{c}}\right\}\right\},
\end{align}
where $\hat{a},~\hat{b},~\hat{c}\in \Z2.$

By taking into account the $\Z2$-grading and the scaling dimension we propose the following Ansatz:
\begin{align}
	\left\{u_{00}(y),\, u_{00}(x)\right\}
	& =a_1u_{00}'(y) \delta(y-x) +a_2u_{00}(y) \delta'(y-x) +a_3\delta'''(y-x)
	,\notag
	\\[2pt]
	\left\{u_{11}(y),\, u_{00}(x)\right\}
	& =a_4u_{11}'(y)\delta(y-x) +a_5u_{11}(y) \delta'(y-x)
	,\notag
	\\[2pt]
	\left\{u_{10}(y),\, u_{00}(x)\right\}
	& =a_6u_{10}'(y)\delta(y-x)+a_7u_{10}(y) \delta'(y-x)
	,\notag
	\\[2pt]
	\left\{u_{01}(y),\, u_{00}(x)\right\}
	& =a_8u_{01}'(y) \delta(y-x)+a_9u_{01}(y) \delta'(y-x)
	,\notag
	\\[2pt]
	\left\{u_{11}(y),\, u_{11}(x)\right\}
	& =b_1u_{00}'(y)\delta(y-x) +b_2u_{00}(y) \delta'(y-x) +b_3\delta'''(y-x)
	,\notag
	\\[2pt]
	\left\{u_{10}(y),\, u_{11}(x)\right\}
	& = b_4u_{01}'(y) \delta(y-x)+b_5u_{01}(y) \delta'(y-x)
	,\notag
	\\[2pt]
	\left\{u_{01}(y),\, u_{11}(x)\right\}
	& = b_6u_{10}'(y) \delta(y-x)+b_7u_{10}(y) \delta'(y-x)
	,\notag
	\\[2pt]
	\left\{u_{10}(y),\, u_{10}(x)\right\}
	& = c_1u_{00}(y)\delta(y-x) +c_2\delta''(y-x)
	,\notag
	\\[2pt]
	\left\{u_{01}(y),\, u_{10}(x)\right\}
	& = c_3u_{11}(y) \delta(y-x)
	,\notag
	\\[2pt]
	\left\{u_{01}(y),\, u_{01}(x)\right\}
	& = d_1u_{00}(y)\delta(y-x) +d_2\delta''(y-x), \label{Ansatz}
\end{align}
where the prime stands for the derivative with respect to $y$ 
and our convention for the delta function is
\begin{align}
	\delta(x-a) &=  \sum_{n \in \mathbb{Z}} e^{in(x-a)},
	\\
	\frac{1}{2\pi}\oint dx \delta(x) &= \frac{1}{2\pi} \int_0^{2\pi} dx \delta(x) = 1.
\end{align}
The constants $a_i,~b_i,~c_i,~d_i$ are fixed by the requirement that \eqref{TransbyGen} and \eqref{Ansatz} reproduce the current transformations \eqref{Utransf1} and \eqref{Utransf2}. 
We note that the Poisson brackets $\{ u_{10}, u_{11} \}, \{ u_{01}, u_{11} \}, \{ u_{10}, u_{10} \} $ and $\{ u_{01}, u_{01} \}$ are symmetric, but all others are antisymmetric. 

For each $\Z2$-graded current, \eqref{TransbyGen} gives the following conditions:
\begin{equation}
	\begin{array}{cc}
		u: & \text{conditions}
		\\ \hline
		u_{00}: &  k_1a_2=k_2a_5=-2i,\qquad k_1a_1=k_2a_4=-i,\qquad k_1a_3 =-\frac{i}{2},
		\\
		 & -k_3a_7=k_4a_9=\frac{3i}{2},\qquad -k_3a_6=k_4a_8 =i
		 \\[8pt]
		 u_{11}: & k_1a_5=k_2b_2=-2i,\qquad k_1a_4=k_2b_1=-i,\qquad k_2b_3 =-\frac{i}{2},
		 \\
		 & -k_3b_5=k_4b_7=\frac{3i}{2},\qquad -k_3b_4=k_4b_6 =i
		 \\[8pt]
		 u_{10}: & k_1a_7=k_2b_5=-\frac{3i}{2},\quad k_1a_6=k_2b_4=-i, \quad 
		 k_3c_1 =k_3c_2=k_4c_3=\frac{1}{2}
		 \\[8pt]
		 u_{01}: & k_1a_9=k_2b_7=-\frac{3i}{2},\quad k_1a_8=k_2b_6=-i, \quad -k_3c_3=k_4d_1=k_4d_2=\frac{1}{2}
	\end{array}
\end{equation}
Solving these conditions give the results:
\begin{alignat}{5}
	k_1 &=i, &\qquad k_2&=i, &\qquad k_3&=i, &\qquad k_4&=-i, & & 
	\notag\\
	a_1& =-1,& a_2&=-2,& a_3&=-\frac12,& a_4&=-1,& \qquad a_5&=-2,
	\notag \\
	a_6 &=-1,& a_7 &=-\frac32,& a_8&=-1,& a_9&=-\frac32,
	\notag \\
	b_1&=-1,& b_2&=-2,& b_3&=-\frac12,& b_4&=-1,& b_5&=-\frac32,
	\notag \\
	b_6&=-1, & b_7&=-\frac32,
	\notag \\
	c_1&=c_2=-\frac{i}{2}, & c_3 &= \frac{i}{2},
	\notag \\
	d_1&=d_2=\frac{i}{2}.
\end{alignat}
Therefore, the $\Z2$-graded currents satisfy the relations
    \begin{align}
	\left\{u_{00}(y),\,u_{00}(x)\right\}
	& =-u_{00}'(y)\delta(y-x) -2u_{00}(y) \delta'(y-x) -\frac{1}{2}\delta'''(y-x)
	,
	\notag\\
	\left\{u_{11}(y),\,u_{00}(x)\right\}
	& =-u_{11}'(y)\delta(y-x) -2u_{11}(y) \delta'(y-x)
	,
	\notag\\
	\left\{u_{10}(y),\,u_{00}(x)\right\}
	& =-u_{10}'(y)\delta(y-x)-\frac{3}{2}u_{10}(y) \delta'(y-x)
	,
	\notag\\
	\left\{u_{01}(y),\,u_{00}(x)\right\}
	& =-u_{01}'(y)\delta(y-x)-\frac{3}{2}u_{01}(y) \delta'(y-x)
	,
	\notag\\
	\left\{u_{11}(y),\,u_{11}(x)\right\}
	& =-u_{00}'(y)\delta(y-x) -2u_{00}(y)\delta'(y-x) -\frac{1}{2}\delta'''(y-x)
	,
	\notag\\
	\left\{u_{10}(y),\,u_{11}(x)\right\}
	& = -u_{01}'(y)\delta(y-x)-\frac{3}{2}u_{01}(y) \delta'(y-x)
	,
	\notag\\
	\left\{u_{01}(y),\,u_{11}(x)\right\}
	& = -u_{10}'(y)\delta(y-x)-\frac{3}{2}u_{10}(y)\delta'(y-x)
	,
	\notag\\
	\left\{u_{10}(y),\,u_{10}(x)\right\}
	& = -\frac{i}{2}u_{00}(y)\delta(y-x) -\frac{i}{2}\delta''(y-x)
	,
	\notag\\
	\left\{u_{01}(y),\,u_{10}(x)\right\}
	& = \frac{i}{2}u_{11}(y)\delta(y-x)
	,
	\notag\\
	\left\{u_{01}(y),\,u_{01}(x)\right\}
	& = \frac{i}{2}u_{00}(y)\delta(y-x) +\frac{i}{2}\delta''(y-x). \label{SuperVirCurAlg}
\end{align}
These relations define a $\Z2$-graded extension of the $\mathcal{N} =1$ super-Virasoro algebra which is recovered by either $ u_{00}, u_{10}$ or $ u_{00}, u_{01}$ subalgebras.

By expanding the $\Z2$-graded currents into modes, we obtain from \eqref{SuperVirCurAlg} the infinite dimensional $\Z2$-graded super-Virasoro algebra equipped with a Poisson-Lie structure. 
As is well known in string theory, fermionic sectors are subject to either periodic (Ramond sector) or antiperiodic (Neveu-Schwarz sector) boundary conditions \cite{Ramond,NS}. 
Assuming, as usual, that the [00]-graded current is periodic, consistency with the current algebra \eqref{SuperVirCurAlg} imposes specific (anti)periodicity conditions on the remaining graded sectors. 
One should note that even the exotic bosonic current $u_{11}$ could satisfy either periodic or antiperiodic boundary condition.  
Therefore, three possible alternatives are admissible in association with 
$
 [11]/[10]/[01] 
$
graded sectors:
\begin{center}
	(i)  R/R/R \qquad (ii) R/NS/NS \qquad (iii) NS/NS/R $\equiv$ NS/R/NS 
\end{center}
where R and NS stand for the Ramond and Neveu-Schwarz sector, respectively. 
We remark that these three admissible boundary conditions already appeared in the string model induced from 1D $\Z2$-graded supersymmetry \cite{aizt}.

Under these periodicity conditions, the $\Z2$-graded currents may be expanded as
\begin{align}
	u_{00}(x)
	& = \sum_{n\in\mathbb Z} L_n e^{inx},
	&
	u_{11}(x)
	& = \sum_{p} H_p e^{ipx},
	\nn
	\\
	u_{10}(x)
	& = \sum_{r} G_r e^{irx},
	&
	u_{01}(x)
	& = \sum_{s} F_s e^{isx} \label{ModeExpansion}
\end{align}
where $ n \in \mathbb{Z}$ is the index for [00]-sector (Ramond sector) and $ p, r, s$ are the  indices for [11], [10], [01]-sectors, respectively. The indices $p, r, s$ take their values in $\mathbb{Z}$ if they are in the Ramond sector or in $  \mathbb{Z}+\frac{1}{2}$ if in the  Neveu-Schwarz sector.
Then we obtain the relations of $\Z2$-graded super-Virasoro algebra:
\begin{align}
	\left\{L_{n},L_{n'}\right\}
	& = i(n'-n)L_{n+n'}+\frac{i}{2}n^{3}\delta_{n+n',\,0},
	\notag \\
	\left\{H_{p},L_{n}\right\}
	& = i(n-p)H_{p+n},
	\notag \\
	\left\{G_{r},L_{n}\right\}
	& = i\left(\frac{n}{2}-r\right)G_{r+n},
	\notag \\
	\left\{F_{s},L_{n}\right\}
	& = i\left(\frac{n}{2}-s\right)F_{s+n},
	\notag \\
	\left\{H_{p},H_{p'}\right\}
	& = i(p'-p)L_{p+p'}+\frac{i}{2}p^{3}\delta_{p+p',\,0},
	\notag \\
	\left\{G_{r},H_{p}\right\}
	& = i\left(\frac{p}{2}-r\right)F_{r+p},
	\notag \\
	\left\{F_{s},H_{p}\right\}
	& = i\left(\frac{p}{2}-s\right)G_{s+p},
	\notag \\
	\left\{G_{r},G_{r'}\right\}
	& = -\frac{i}{2}L_{r+r'}+\frac{i}{2}r^{2}\delta_{r+r',\,0},
	\notag \\
	\left\{F_{s},G_{r}\right\}
	& = \frac{i}{2}H_{s+r},
	\notag \\
	\left\{F_{s},F_{s'}\right\}
	& = \frac{i}{2}L_{s+s'}-\frac{i}{2}s^{2}\delta_{s+s',\,0} \label{SuperVirRelations}
\end{align}
where the indices take the following values according to the boundary conditions. 
(i) all the indices are integer, (ii) $ n, n', p, p' \in \mathbb{Z}$ and $ r, r', s, s' \in \mathbb{Z}+\frac{1}{2},$ (iii) $ n, n', s, s' \in \mathbb{Z}$ and $ p, p', r, r' \in \mathbb{Z}+ \frac{1}{2}. $

We derive the Lie-Poisson algebra with a [00]-graded central extension, in which the central charge takes a fixed value. 
{\color{mycol}
The factors of the central terms $n^3$ and $r^2$ for [00]-graded and [10]-graded currents in \eqref{SuperVirRelations} differ from the standard choice in CFT (note that the same applies to the [11]- and [01]-graded currents). This discrepancy arises from our convention for the mode expansion \eqref{ModeExpansion}, which is defined on $S^1,$  instead of the punctured complex plane $\mathbb{C}^*=\mathbb{C}-\{0\}$, as customary
in CFT, where the mode expansion is typically given by  $ u_{00}(z) = \sum_{n \in \mathbb{Z}} L_n z^{-n-2}$.
We recall that the central charges proportional to $n^3, r^2$ and $ n(n^2-1), r^2-\frac{1}{4}$ (as derived in CFT)  differ by a trivial cocycle which can be reabsorbed in the redefinitions of the generators.
}
The most general $\Z2$-graded extension of the super-Virasoro algebra admits two central extensions: one [00]-graded and the other [11]-graded \cite{NAJS}. 
However, the extension admitting a [11]-graded central charge is not obtained within the present framework. 
Another open problem is the possible isomorphism between two algebras characterized by R/R/R and R/NS/NS boundary conditions. It is known that the $\mathcal{N} = 2 $ Ramond and Neveu-Schwarz superalgebras are isomorphic  \cite{schsei}. A similar structure appears in the $\mathcal{N} = 2 $ 
super-Virasoro algebra and the $\Z2$-graded super-Virasoro algebra derived in this section, as the latter also possesses two fermionic currents: one [10]-graded and the other [01]-graded. 
Therefore, it is important to clarify whether the isomorphism exists.

Finally, we comment that a  different type of $\Z2$-graded extension of super-Virasoro algebra is discussed in \cite{zhe} where the $(00,11,10,01)$-graded currents  have the scaling dimensions $(2,1,\frac{3}{2},\frac32), $ respectively. In contrast, the $\mathbb{Z}_2$-graded super-Virasoro algebra presented in this work features scaling dimensions $(2, 2, \tfrac{3}{2}, \tfrac{3}{2})$ and is associated with the $\Z2$-super-Liouville equation \eqref{SLsol}.\par
The possibility of having inequivalent ${\mathbb Z}_2^2$-graded extensions of superalgebras was already mentioned in Introduction. 
Another illustrative example is provided by the two-dimensional superPoincar\'e algebra, which  admits ${\mathbb Z}_2^2$-graded extensions with
two $[11]$-graded translations \cite{tol}, two $[00]$-graded translations \cite{brusigma} and one $[00]$-graded plus one $[11]$-graded translation \cite{aizt}.

\section{Zero-curvature formulation of the $\Z2$-graded super-Liouville equation} \label{SEC:ZCF}
\setcounter{equation}{0}

In this section and the following one, we investigate the integrability of the  $\Z2$-super-Liouville equation derived via the soldering procedure described in \S \ref{SEC:Equatoion}. 
We first show that the $\Z2$-super-Liouville equation can be formulated within the zero-curvature framework. Three inequivalent Lax operators are found to be admissible for this formulation. Then, we explicitly construct a solution to the equation.

\subsection{Lax operators and zero-curvature equation}

We return to the superfields defined on superspace with [00] and [10]-graded coordinates. 
We now adopt a notation more suitable for the zero-curvature framework. 
The coordinate of superspace are denoted by
\begin{equation}
	[00] \ x_+, \ x_-, \qquad [10] \ \theta_+, \ \theta_-
\end{equation}
and we write the corresponding covariant derivatives as
\begin{equation}
	D_{\pm} = \partial_{\theta_{\pm}} + i \theta_{\pm} \partial_{\pm}, \qquad \partial_{\pm} := \partial_{x_{\pm}}. 
\end{equation}
They satisfy the following relations:
\begin{align}
	\{ D_{\pm}, D_{\pm} \} &= 2i  \partial_{\pm}, \qquad \{ D_+, D_- \}  = 0. 
\end{align}
Denoting the superfields by  $\Phi_{00}(x_{\pm}, \theta_{\pm})$ and $ \Phi_{11}(x_{\pm}, \theta_{\pm})$ we define the $[10]$-graded Lax operators
\begin{align}
		\TL_{\pm} &= \mp D_{\pm} \Phi + e^{\pm \Phi} P_{\pm} e^{\mp\Phi} = \mp D_{\pm} \Phi + A_{00} P_{\pm} + i A_{11} Q_{\pm}, 
	\label{LaxOps} \\
	\Phi &= \frac{1}{2}(\Phi_{00} K_0 + \Phi_{11} L_0)
\end{align}
where 
\begin{equation}
	A_{00} = e^{\frac{1}{2}\Phi_{00}} \cosh \frac{1}{2} \Phi_{11},
	\qquad 
	A_{11} = e^{\frac{1}{2}\Phi_{00}} \sinh \frac{1}{2} \Phi_{11}. 
	\label{DefA}
\end{equation}
Following the general construction of integrable systems, we consider the linear system
\begin{equation}
	(D_{\pm} - \TL_{\pm}) T = 0 
\end{equation}
where $ T $ is an element of the group generated by $\g. $ 
The compatibility condition of the system is the zero-curvature condition which is in the present case given by
\begin{equation}
	D_+ \TL_- + D_- \TL_+ - \{ \TL_+, \TL_- \} = 0.   \label{ZeroCurv}
\end{equation}
Using only the algebraic structure of $\g$, one may obtain the following equations from the zero-curvature condition:
\begin{alignat}{2}
	D_+ D_- \Phi_{00} &= e^{\Phi_{00}} \cosh \Phi_{11},
	& \qquad
	D_+ D_- \Phi_{11} &= e^{\Phi_{00}} \sinh \Phi_{11} 
	\label{Z22SLeq}
\end{alignat}
and 
\begin{align}
	D_{\pm} A_{00} &= \frac{1}{2}( D_{\pm} \Phi_{00} \cdot A_{00} + D_{\pm} \Phi_{11}\cdot A_{11} ),
	\notag \\
	D_{\pm} A_{11} &= \frac{1}{2}( D_{\pm} \Phi_{00} \cdot A_{11} + D_{\pm} \Phi_{11}\cdot A_{00} ).
	\label{A-equations}
\end{align}
As is readily seen from \eqref{DefA}, the equations in \eqref{A-equations} are identities. 
Whereas the equations in \eqref{Z22SLeq} are non-trivial dynamical equations which are identical to \eqref{SLsol}.

\subsection{Alternative Lax operators}

We show that the $\Z2$-super-Liouville equation in the component form \eqref{CompEq1}, \eqref{CompEq2} admits an alternative Lax operator formulation. 
Let us  consider a $\Z2$-graded superspace with coordinates
\begin{equation}
	[00] \ x_+, \ x_-, \qquad [10] \ \theta_{10}, \qquad [01]\ \theta_{01} \label{Z2SPcoordinates}
\end{equation}
where $ \theta_{10}, \theta_{01}$ are nilpotent and mutually commuting. 
The corresponding covariant derivatives are defined as
\begin{equation}
	D_{10} = \partial_{\theta_{10}} + i \theta_{10} \partial_+,
	\qquad
	D_{01} = \partial_{\theta_{01}} + i \theta_{01} \partial_-
\end{equation}
and satisfy the relations:
\begin{equation}
	\{ D_{10}, D_{10} \} = 2i\partial_+, \qquad 
	\{ D_{01}, D_{01} \} = 2i\partial_-, \qquad
	[D_{10}, D_{01}] = 0.
\end{equation}

We introduce the [00]-graded superfield 
$ \tilde{\Phi}(x_{\pm}, \theta_{10}, \theta_{01})$ defined by
\begin{equation}
	\tilde{\Phi} = \frac{1}{2} (\tilde{\Phi}_{00} K_0 + \tilde{\Phi}_{11} L_0)
\end{equation}
where $ \tilde{\Phi}_{00} $ and $ \tilde{\Phi}_{11}$ are also superfields on the $\Z2$-graded superspace \eqref{Z2SPcoordinates}. 
We define the [10] and [01]-graded Lax operators:
\begin{align}
	\mathcal{L}_{10} &= -D_{10} \tilde{\Phi} + e^{\tilde{\Phi}} P_+ e^{-\tilde{\Phi}} = -D_{10} \tilde{\Phi} + \tilde{A}_{00} P_+ + i\tilde{A}_{11} Q_+,
	\notag \\
	\mathcal{L}_{01}&= D_{01} \tilde{\Phi} + e^{-\tilde{\Phi}} P_+ e^{\tilde{\Phi}} = D_{10} \tilde{\Phi} + \tilde{A}_{00} Q_- - i\tilde{A}_{11} P_-
\end{align}
where 
\begin{equation}
	\tilde{A}_{00} = e^{\frac{1}{2}\tilde{\Phi}_{00}} \cosh \frac{1}{2} \tilde{\Phi}_{11},
	\qquad 
	\tilde{A}_{11} = e^{\frac{1}{2}\tilde{\Phi}_{00}} \sinh \frac{1}{2} \tilde{\Phi}_{11}. 
	\label{DeftildeA}
\end{equation}

The zero-curvature condition for these Lax operators takes the form
\begin{equation}
	D_{10} \mathcal{L}_{01} - D_{01} \mathcal{L}_{10} - [\mathcal{L}_{10},\mathcal{L}_{01}] = 0
\end{equation}
Using only the algebraic structure of $\g$ as in the previous subsection, one may obtain the following equations from the zero-curvature condition:
\begin{equation}
	D_{10} D_{01} \tilde{\Phi}_{00} = i e^{\tilde{\Phi}_{00}} \sinh \tilde{\Phi}_{11}, 
	\qquad 
	D_{10} D_{01} \tilde{\Phi}_{11} = i e^{\tilde{\Phi}_{00}} \cosh \tilde{\Phi}_{11} 
	\label{SLalternative}
\end{equation}
The difference from the previous choice of Lax operators is that no equations corresponding to \eqref{A-equations} are obtained in this case. 

By expanding the superfields $ \tilde{\Phi}_{00}$ and $ \tilde{\Phi}_{11}$ as
\begin{align}
	\tilde{\Phi}_{00} &= \varphi_{00} + \theta_{10} \psi_{10} + i \theta_{01} \bar{\psi}_{01} - i \theta_{10} \theta_{01} F_{11},
	\notag \\
	\tilde{\Phi}_{11} &= \varphi_{11} + \theta_{10} \psi_{01} + i \theta_{01} \bar{\psi}_{10} - i \theta_{10} \theta_{01} F_{00},	
\end{align}
one readily verifies that the equations \eqref{SLalternative} are equivalent to \eqref{CompEq1} and \eqref{CompEq2}.

\subsection{Lax operators with spectral parameter}

It is possible to formulate the component equations (see \eqref{CompEq1}, \eqref{CompEq2} and  \eqref{ComponentEq2}) by the Lax operators with a spectral parameter and without using the superfields. Let us introduce the loop extension of $\g$ defined by
\begin{align}
	K_n &:= \lambda^n \otimes K_0, \qquad K^{\pm}_n := \lambda^n \otimes K_{\pm}, \qquad 
	L_n :=\lambda^n \otimes L_0, \qquad L^{\pm}_n := \lambda^n \otimes L_{\pm},
	\notag \\
	P^{\pm}_n &:= \lambda^n \otimes P_{\pm}, \qquad Q^{\pm}_n := \lambda^n \otimes Q_{\pm}. 
	\label{LoopExt1}
\end{align}
where $ n \in \mathbb{Z} $ and $ \lambda $ is a parameter. These form an infinite dimensional $\Z2$-graded Lie superalgebra. This algebra has central extensions and derivations with non-trivial grading \cite{NAJS}. However, these are not necessary for our current purpose. 

We define the $[00]$-graded Lax operators using the elements of \eqref{LoopExt1}:
\begin{align}
	\mathscr{L}_+ &= -\partial_+ \varphi_{00} K_0 -\partial_+ \varphi_{11} L_0 -i(K^+_2 + K^-_2) + \bar{\psi}_{10} P^+_1 + i \bar{\psi}_{01} Q^+_1,
	\notag \\[3pt]
	\mathscr{L}_- &= ie^{2\varphi_{00}} (\cosh 2\varphi_{11} \cdot K^-_{-2} +  \sinh 2\varphi_{11} \cdot L^-_{-2}) + e^{\varphi_{00}} (\Lambda_{10} P^-_{-1} + i \Lambda_{01} Q^-_{-1}).
\end{align}
It is then straightforward to verify that the zero-curvature condition
\begin{equation}  
	\partial_- \mathscr{L}_+ - \partial_{+} \mathscr{L}_- + [\mathscr{L}_+, \mathscr{L}_-]
	= 0
\end{equation}
gives the component form of the $\Z2$-super-Liouville equation. 

 To show the spectral parameter dependence more explicitly, we present the matrix representation of $\mathscr{L}_{\pm} $ in the space of six-dimensional representation of $\g$ given in \S \ref{SubSec:Rep}: 
\begin{equation}
	\mathscr{L}_+ = 
	\left[
	\begin{array}{cccc|cc}
		-\partial_{+} \varphi_{00} & -i\lambda^2 & -\partial_{+} \varphi_{11} & 0 & \lambda \bar{\psi}_{10} & i \lambda \bar{\psi}_{01}
		\\[3pt]
		-i\lambda^2 & \partial_{+} \varphi_{00} & 0 & \partial_{+} \varphi_{11} & 0 & 0
		\\[3pt]
		-\partial_{+} \varphi_{11} & 0 & -\partial_{+} \varphi_{00} & -i\lambda^2 & -\lambda \bar{\psi}_{01} & -i \lambda \bar{\psi}_{10}
		\\[3pt] 
		0 & \partial_{+} \varphi_{11} & -i\lambda^2 & \partial_{+} \varphi_{00} & 0 & 0 
		\\[3pt]\hline
		0 & -\lambda \bar{\psi}_{10} & 0 & -\lambda \bar{\psi}_{01} & 0 & 0
		\\[3pt]
		0 & -i\lambda \bar{\psi}_{01} & 0 & -i\lambda \bar{\psi}_{10} & 0 & 0
	\end{array}
	\right],
\end{equation}
\begin{equation}
	\mathscr{L}_- =
	\left[
	\begin{array}{cccc|cc}
		0 & 0 & 0 & 0 & 0 & 0
		\\[3pt]
		i\lambda^{-2} f_{00} & 0 & i\lambda^{-2} f_{11} & 0 & -\lambda^{-1} e^{\varphi_{00}}\Lambda_{10} & -i\lambda^{-1} e^{\varphi_{00}}\Lambda_{01}
		\\[3pt]
		0 & 0 & 0 & 0 & 0 & 0
		\\[3pt] 
		i\lambda^{-2} f_{11} & 0 & i\lambda^{-2} f_{00} & 0 & \lambda^{-1} e^{\varphi_{00}}\Lambda_{01} & i\lambda^{-1} e^{\varphi_{00}}\Lambda_{10}
		\\[3pt]\hline
		-\lambda^{-1} e^{\varphi_{00}} \Lambda_{10} & 0 & -\lambda^{-1} e^{\varphi_{00}} \Lambda_{01} & 0 & 0 & 0
		\\[3pt]
		-i\lambda^{-1} e^{\varphi_{00}} \Lambda_{01} & 0 & -i\lambda^{-1} e^{\varphi_{00}} \Lambda_{10} & 0 & 0 & 0
	\end{array}
	\right]
\end{equation}
where $ f_{00} := e^{2\varphi_{00}} \cosh 2\varphi_{11}, f_{11} := e^{2\varphi_{00}} \sinh 2\varphi_{11}. $ 


\subsection{Reconstruction of the superfields}\label{SEC:LSA}

In this subsection, we solve the $\Z2$-super-Liouville equation \eqref{Z22SLeq} by the method developed in \cite{LezSav1,LezSav2,LezSav3}. 
First, note that solutions of the zero curvature condition is pure gauge: 
\begin{equation}
	\TL_{\pm} = (D_{\pm}T) T^{-1}. \label{PureGauge}
\end{equation}
We write $T$ in two different ways (generalized Gauss decomposition)
\begin{equation}
	T = e^{\mp \Phi} N_{\pm} B_{\mp}, \quad N_{\pm} \in \exp(\g_{\pm}), \ B_{\mp} \in \exp(\h \oplus \g_{\mp})
	\label{Tdecomposition}
\end{equation}
Substituting this into \eqref{PureGauge} and comparing with \eqref{LaxOps}, we obtain
\begin{equation}
	N_{\pm} (D_{\pm} B_{\mp}) B_{\mp}^{-1} N_{\pm}^{-1}  =  - D_{\pm} N_{\pm} \cdot N_{\pm}^{-1} + e^{\pm 2 \Phi} P_{\pm} e^{\mp 2\Phi}.
\end{equation}
The RHS is in $\g_{\pm}$, while the LHS is in $\h \oplus \g_{\mp}$ which implies the followings
\begin{align}
	D_{\pm} B_{\mp} &= 0, 
	\label{chiralB} \\
	(D_{\pm} N_{\pm}) N_{\pm}^{-1} &= e^{\pm 2\Phi} P_{\pm} e^{\mp 2\Phi} = e^{\Phi_{00}} (\cosh \Phi_{11}\cdot P_{\pm} + i \sinh \Phi_{11} \cdot Q_{\pm} ). 
	\label{Const1}
\end{align}
On the other hand, substitution $ T = e^{\pm \Phi} N_{\mp}B_{\pm} $ into \eqref{PureGauge} gives
\begin{align}
		(D_{\pm} B_{\pm}) B_{\pm}^{-1} &= - N_{\mp}^{-1} D_{\pm} N_{\mp} 
	+ N_{\mp}^{-1} (\mp 2 D_{\pm} \Phi + P_{\pm} ) N_{\mp}
	\nn \\
	&= P_{\pm} + (\text{terms in } \h \oplus \g_{\mp}).
	\label{Const2}	
\end{align}

 Now we parametrize $B_{\pm}$ as 
\begin{align}
B_{\pm} &= e^{f_{\pm} K_0} e^{g_{\pm} L_0} e^{q_{\pm} K_{\pm}} e^{r_{\pm} L_{\pm}} e^{\alpha_{\pm} P_{\pm}} e^{\beta_{\pm} Q_{\pm}} 
\label{Bparam}
\end{align}
where the parameters are superfields (functions of $x_{\pm}, \theta_{\pm}$) with the following $\Z2$-gradings
\begin{equation}
	[00] \ f_{\pm},\ q_{\pm}, \quad [11]\ g_{\pm}, \ r_{\pm},  \quad 
	[10] \ \alpha_{\pm}, \quad [01] \ \beta_{\pm}
\end{equation}
and are $\Z2$-commutative. 
The condition \eqref{chiralB} implies that all the superfields are chiral, i.e., 
$ f_+(x_+,\theta_+), \ f_-(x_-, \theta_-),$ etc. 
From \eqref{Const2} we obtain several constraints on the superfields:
\begin{align}
	D_{\pm} q_{\pm} \pm (D_{\pm} \alpha_{\pm}) \alpha_{\pm} \pm (D_{\pm} \beta_{\pm}) \beta_{\pm} &= 0,
	\notag \\
	D_{\pm} r_{\pm} \pm 2i (D_{\pm}\beta_{\pm}) \alpha_{\pm} &= 0
	\label{Rel5}
\end{align}
and
	\begin{equation}
	D_{\pm} \alpha_{\pm} = e^{\mp f_{\pm}} \cosh g_{\pm}, \qquad
	D_{\pm} \beta_{\pm} = \mp i e^{\mp f_{\pm}} \sinh g_{\pm}.
	\label{Rel6}
\end{equation}

Let $\ket{\ }$ be a lowest weight vector introduced in \S \ref{SubSec:Rep}. 
The following relations follow immediately from the lowest weight nature of $\ket{\ }:$
\begin{align}
	\bra{\ } e^{\Phi} T = \bra{\ } B_-, \qquad (e^{-\Phi}T)^{-1} \ket{\ } =  B_+^{-1} \ket{\ }
\end{align}
which give the reconstruction of $\Phi_{00}$ and $\Phi_{11}$ in terms of the chiral superfields in \eqref{Bparam}:
\begin{equation}
	\bra{\ } e^{2\Phi} \ket{\ } = \bra{\ } B_-(x_-,\theta_-) B_+^{-1}(x_+,\theta_+) \ket{\ }. \label{Reconstruction}
\end{equation}
The LHS is easily computed to get
\begin{equation}
	\braket{00 | e^{2\Phi} |00} = e^{-\Phi_{00}} \cosh \Phi_{11},
	\qquad 
	\braket{11|e^{2\Phi}|00} = -e^{-\Phi_{00}} \sinh \Phi_{11}.
\end{equation}
After a lengthy computation, the corresponding RHS can be obtained, providing the solution to the $\Z2$-super-Liouville equation: 
\begin{align}
	e^{-\Phi_{00}} \cosh \Phi_{11} &= e^{f_+ - f_-} \big( W_{00} \cosh(g_+ -g_-) + W_{11} \sinh (g_+-g_-)  \big),
	\label{Solution1} \\
	e^{-\Phi_{00}} \sinh \Phi_{11} &= -e^{f_+ - f_-} 
	\big(  W_{00}\sinh(g_+-g_-) + W_{11} \cosh(g_+-g_-) \big)
	\label{Solution2}
\end{align}
where 
\begin{align}
	W_{00} &= 1+\alpha_+ \alpha_- + \beta_+ \beta_- - q_+ q_- -R_+ R_-,
	\notag\\
	W_{11} &= - i \alpha_+ \beta_- + i \alpha_- \beta_+ - q_+ R_- - q_- R_+
\end{align}
with $ R_{\pm} = r_{\pm} \pm i \alpha_{\pm} \beta_{\pm}.$

\subsection{Checking the solution}

We here present a proof that \eqref{Solution1} and \eqref{Solution2} are indeed the solutions to the $\Z2$-super-Liouville equation.  We do this by direct computation. 
Solving \eqref{Solution1} and \eqref{Solution2} for $ \cosh \Phi_{11} $ and $\sinh \Phi_{11}$ and using $\cosh^2 x - \sinh^2 x = 1, $ we get
\begin{align}
	e^{-2\Phi_{00}} &= e^{2(f_+-f_-)} (W_{00}^2 - W_{11}^2),
	\label{PhiWW} \\
	\tanh \Phi_{11} &= - \frac{ W_{00}\sinh(g_+-g_-) + W_{11} \cosh(g_+-g_-)}{ W_{00} \cosh(g_+ -g_-) + W_{11} \sinh (g_+-g_-)}.
\end{align}
Then, it is not difficult to compute the covariant derivatives of the superfields:
\begin{align}
	D_+ D_- \Phi_{00} &= \frac{1}{(W_{00}^2-W_{11}^2)^2} 
	\big[  (W_{00}^2 + W_{11}^2) (D_+ W_{00}\cdot D_- W_{00} - D_+ W_{11}\cdot D_- W_{11})
	\notag \\
	&-2 W_{00} W_{11} (D_+ W_{00}\cdot D_- W_{11} - D_- W_{00}\cdot D_+ W_{11}) 
	\notag \\
	&- (W_{00}^2 - W_{11}^2) (W_{00} D_+ D_-W_{00} - W_{11} D_+ D_-W_{11}) \big]
	\label{DDPhi00}
\end{align}
and
\begin{align}
	D_+ D_- \Phi_{11} &= \frac{1}{(W_{00}^2-W_{11}^2)^2} 
	\big[  (W_{00}^2 + W_{11}^2) (D_+ W_{00}\cdot D_- W_{11} - D_- W_{00}\cdot D_+ W_{11})
	\notag \\
	&-2 W_{00} W_{11} (D_+ W_{00}\cdot D_- W_{00} - D_+ W_{11}\cdot D_- W_{11}) 
	\notag \\
	&- (W_{00}^2 - W_{11}^2) (W_{00} D_+ D_-W_{11} - W_{11} D_+ D_-W_{00}) \big].
	\label{DDPhi11}
\end{align}

We rewrite the above equations by using the constraints \eqref{Rel5} and \eqref{Rel6}. 
Using the constraints \eqref{Rel5}, one may see that
\begin{align}
	D_{\pm} W_{00} &= \pm \mathscr{A}_{\mp} D_{\pm} \alpha_{\pm} \mp \mathscr{B}_{\mp} D_{\pm} \beta_{\pm},
	\notag \\
	D_{\pm} W_{11} &= \mp i (\mathscr{B}_{\mp} D_{\pm} \alpha_{\pm} + \mathscr{A}_{\mp} D_{\pm} \beta_{\pm} )
\end{align}
where
\begin{align}
	\mathscr{A}_{\pm} &= \alpha_{\pm} + q_{\pm} \alpha_{\mp} + i R_{\pm} \beta_{\mp},
	\quad \deg(\mathscr{A}_{\pm}) = [10],
	\notag\\
	\mathscr{B}_{\pm} &= \beta_{\pm} + q_{\pm} \beta_{\mp} -i R_{\pm} \alpha_{\mp}, 
	\quad \deg(\mathscr{B}_{\pm}) = [01].
\end{align}
It follows that
\begin{align}
	D_+ W_{00}\cdot D_- W_{00} - D_+ W_{11}\cdot D_- W_{11} &= 
	(\Ap \Am + \Bp \Bm) \mathcal{D}_{00} + (\Am \Bp - \Ap \Bm) \mathcal{D}_{11},
	\notag\\
	D_+ W_{00}\cdot D_- W_{11} - D_+ W_{11}\cdot D_- W_{00} &= 
	-i(\Ap \Am + \Bp \Bm) \mathcal{D}_{11} + i(\Am \Bp - \Ap \Bm) \mathcal{D}_{00}
\end{align}
where
\begin{align}
	\mathcal{D}_{00} &:= D_+ \alpha_+\cdot D_-\alpha_- - D_+ \beta_+\cdot D_- \beta_- \stackrel{\eqref{Rel6}}{=} e^{-f_+ + f_-} \cosh(g_+ - g_-),
	\notag \\
	\mathcal{D}_{11} &:= D_+ \alpha_+\cdot D_-\beta_- + D_-\alpha_-\cdot D_+ \beta_+\stackrel{\eqref{Rel6}}{=} -ie^{-f_+ + f_-} \sinh(g_+ - g_-)
	\label{D00D11}
\end{align}
and the following relations hold true:
\begin{align}
	\Ap \Am + \Bp \Bm 
	&= 
	(\alpha_+\alpha_- + \beta_+\beta_-) W_{00} + i (\alpha_-\beta_+ - \alpha_+ \beta_-) W_{11},
	\notag \\
	\Am \Bp - \Ap \Bm 
	&= 
	-i (\alpha_+\alpha_- + \beta_+\beta_-) W_{11} + (\alpha_-\beta_+ - \alpha_+ \beta_-) W_{00}.
\end{align}
Then, we obtain the second order derivatives:
\begin{align}
	D_+ D_- W_{00} &= -(1-\alpha_+\alpha_- - \beta_+\beta_-) \mathcal{D}_{00}
	+(\alpha_-\beta_+ -\alpha_+ \beta_-) \mathcal{D}_{11}
	\notag\\
	D_+ D_- W_{11} &= i(1-\alpha_+\alpha_- - \beta_+\beta_-) \mathcal{D}_{11} 
	+i(\alpha_-\beta_+ -\alpha_+ \beta_-)\mathcal{D}_{00}.
\end{align}
With these, one may compute \eqref{DDPhi00} and \eqref{DDPhi11} as follows:
\begin{align}
	D_+ D_- \Phi_{00} &= \frac{1}{W_{00}^2-W_{11}^2} (W_{00} \mathcal{D}_{00} + i W_{11} \mathcal{D}_{11})
	\notag\\
	&\stackrel{\eqref{D00D11}}{=} \frac{e^{-f_+ + f_-}}{W_{00}^2 - W_{11}^2} \big(W_{00}\cosh(g_+-g_-) + W_{11} \sinh(g_+-g_-) \big)
	\notag\\
	&= e^{\Phi_{00}} \cosh\Phi_{11}
\end{align}
and
\begin{align}
	D_+ D_- \Phi_{11} &= \frac{-1}{W_{00}^2-W_{11}^2} (W_{11} \mathcal{D}_{00} + i W_{00} \mathcal{D}_{11})
	\notag\\
	&\stackrel{\eqref{D00D11}}{=} \frac{-e^{-f_+ + f_-}}{W_{00}^2 - W_{11}^2} \big(W_{11}\cosh(g_+-g_-) + W_{00} \sinh(g_+-g_-) \big)
	\notag\\
	&= e^{\Phi_{00}} \cosh\Phi_{11}.
\end{align}
This completes the check of the solutions.

\section{$\Z2$-graded B\"acklund transformations} \label{SEC:Backlund}
\setcounter{equation}{0}

It is known that the Liouville equation has two B\"acklund transformations, one is the transformation to the free equation and the other is the auto-B\"acklund transformation. 
These play important roles of integrability of the equation and the corresponding transformations for the super-Liouville equation is also known \cite{ChaiKu}. 
In this section, we present a $\Z2$-graded version of those transformations. 

Suppose that $\Phi_{00} $ and $\Phi_{11}$ solve the $\Z2$-super-Liouville equation \eqref{Z22SLeq}. 
We denote the transformed superfields  by $\tilde{\Phi}_{00}, \tilde{\Phi}_{11}$ and write the linear combination as
\begin{equation}
	V_{\pm} := \frac{1}{2}(\Phi_{00} \pm \tilde{\Phi}_{00}), \qquad
	W_{\pm} := \frac{1}{2}(\Phi_{11} \pm \tilde{\Phi}_{11}). \label{VWdef}
\end{equation}
To define B\"acklund transformations we need introduce two additional superfields:
\begin{equation}
	\Lambda(x_{\pm}, \theta_{\pm}), \quad \Gamma(x_{\pm}, \theta_{\pm}), \qquad 
	\deg(\Lambda) = [10], \quad \deg(\Gamma) = [01].
\end{equation}

Then, a B\"acklund transformation is defined as follows
\begin{align}
	D_+ V_+ &= \frac{a}{2} e^{V_-}(\Lambda \cosh W_- +\Gamma \sinh W_-),
	\notag\\
	D_- V_- &= \frac{1}{2} e^{V_+}(\Lambda \cosh W_+ +\Gamma \sinh W_+)
	\notag\\
	D_+ W_+ &= \frac{a}{2} e^{V_-}(\Lambda \sinh W_- +\Gamma \cosh W_-),
	\notag\\
	D_- W_- &= \frac{1}{2} e^{V_+}(\Lambda \sinh W_+ +\Gamma \cosh W_+)
	\notag\\
	D_+ \Lambda &= e^{V_-} \cosh W_-
	,\quad
	D_- \Lambda = -\frac{1}{a}e^{V_+} \cosh W_+
	\notag\\ 
	D_+ \Gamma &= e^{V_-} \sinh W_-
	,\quad
	D_- \Gamma = -\frac{1}{a}e^{V_+} \sinh W_+
	, \label{Backlund}
\end{align}
where $ a $ is a non-zero constant. 
We claim that  $\tilde{\Phi}_{00}, \tilde{\Phi}_{11}$ solve the free equations: 
\begin{equation}
	D_+ D_- \tilde{\Phi}_{00} = D_+ D_- \tilde{\Phi}_{11} = 0. \label{FreeEq}
\end{equation}
This can be verified straightforwardly.  
The relations \eqref{Backlund} imply
\begin{align}
	2D_+ D_- V_{\pm} = e^{\Phi_{00}} \cosh \Phi_{11}, \qquad 
	2D_+ D_- W_{\pm} = e^{\Phi_{00}} \sinh \Phi_{11}.
\end{align} 
Assuming that $\Phi_{00}$ and $ \Phi_{11}$ solve the $\Z2$-super-Liouville equation \eqref{Z22SLeq}, these relations lead directly to the free equations \eqref{FreeEq}. 

An auto-B\"acklund transformation, which transforms the $\Z2$-super-Liouville equation to themselves, is defined by
\begin{align}
	D_+ V_+ &= a\big( \Lambda \cosh V_- \cosh W_- +\Gamma \sinh V_- \sinh W_-\big),
	\notag \\
	D_- V_- &= e^{V_+} (\Lambda \cosh W_+ +\Gamma \sinh W_+),
	\notag \\
	D_+ W_+ &= a\big(\Lambda \sinh V_- \sinh W_- +\Gamma \cosh V_- \cosh W_-\big),
	\notag \\
	D_- W_- &= e^{V_+} (\Lambda \sinh W_+ +\Gamma \cosh W_+),
	\notag \\
	D_+ \Lambda &= \sinh V_- \cosh W_-, \qquad 
	D_- \Lambda = -\frac{1}{a} e^{V_+} \cosh W_+,
	\notag \\
	D_+ \Gamma &= \cosh V_- \sinh W_-, \qquad 
	D_- \Gamma = -\frac{1}{a} e^{V_+} \sinh W_+,
	\label{AutoBack}
\end{align}
where $ a $ is a non-zero constant {\color{mycol} and $V_{\pm}, W_{\pm}$ are given by \eqref{VWdef}.} 
{\color{mycol} 
It is easy to verify that if $\Phi_{00}$ and $ \Phi_{11}$ satisfy \eqref{Z22SLeq},  then  $\tilde{\Phi}_{00}$ and $ \tilde{\Phi}_{11}$  do so as well. 
We briefly outline the computation. Using the relations in \eqref{AutoBack}, one may obtain the followings:
\begin{align}
	D_+ D_- V_+ &= e^{V_+} (\cosh V_- \cosh W_+ \cosh W_- + \sinh V_- \sinh W_+ \sinh W_-),
	\notag \\
	D_+ D_- V_- &= e^{V_+} (\cosh V_- \sinh W_+ \sinh W_- + \sinh V_- \cosh W_+ \cosh W_-),
	\notag \\
	D_+ D_- W_+ &= e^{V_+} (\sinh V_- \cosh W_+ \sinh W_- + \cosh V_- \sinh W_+ \cosh W_-),
	\notag \\
	D_+ D_- W_- &= e^{V_+} (\sinh V_- \sinh W_+ \cosh W_- + \cosh V_- \cosh W_+ \sinh W_-).
\end{align}
It follows that
\begin{align}
	D_+ D_- (V_+ \pm V_-) &= e^{V+ \pm V_-} \cosh(W_+ \pm W_-),
	\notag\\
	D_+ D_- (W_+ \pm W_-) &= e^{V+ \pm V_-} \sinh(W_+ \pm W_-).
\end{align}
These are the $\Z2$-super-Liouville equations for $ \Phi_{00}, \Phi_{11}$ and $ \tilde{\Phi}_{00}, \tilde{\Phi}_{11}$.
}

The generalized current conservation laws arise from the (auto-)B\"acklund transformations. 
We define the following $\Z2$-graded currents:
\begin{alignat}{2}
	J_{00}^{\pm} &= D_{\mp}\Lambda, &  
	J_{11}^{\pm} &= D_{\mp} \Gamma,
	\nn\\
	J_{10}^{\pm} &= \pm J_{00}^{\pm} \Lambda \mp J_{11}^{\pm} \Gamma, 
	& \qquad 
	J_{01}^{\pm} &= \pm J_{00}^{\pm} \Gamma \mp J_{11}^{\pm} \Lambda.
\end{alignat}
Here $J_{00}^{\pm}$ and $J_{11}^{\pm}$ are given by \eqref{Backlund} for the Bäcklund transformation and by \eqref{AutoBack} for the auto-Bäcklund transformation.
These currents satisfy the generalized conservation laws
\begin{equation}
	D_+ J^+_{\alpha} + D_- J^-_{\alpha} = 0, \quad \alpha \in \{ 00, 11, 01, 10 \}
\end{equation}
which can be readily verified by using the relations
\begin{align}
	(D_{\pm} J_{00}^{\pm}) \Lambda - (D_{\pm} J_{11}^{\pm}) \Gamma = 0,
	\nn \\
	(D_{\pm} J_{00}^{\pm}) \Gamma - (D_{\pm} J_{11}^{\pm}) \Lambda = 0.
\end{align}

{\color{mycol}
For a given solution of either the free equation or the $\Z2$-super-Liouville equation, one can use the Bäcklund transformations \eqref{Backlund} or \eqref{AutoBack} to generate another solution of the $\Z2$-super-Liouville equation. 
It is also expected that the Bianchi's permutability theorem, which allows one to combine two solutions to obtain a third, can be extended to the $\Z2$-setting.  
However, verifying these properties is beyond the scope of the present work and is left for future investigation. 
The reason is that the involved computations are cumbersome and highly nontrivial due to the presence of additional superfields $ \Lambda, \Gamma  $ which satisfy nontrivial relations with $\Phi_{00}$ and $\Phi_{11}$. 
We point out that the alternative method described in \S \ref{SEC:LSA} is much simpler and more transparent approach to  solving the $\mathbb{Z}_2$-graded equation.
}

\section{Conclusions}
\setcounter{equation}{0}

We have derived an integrable $\Z2$-graded  extension of the super-Liouville equation and investigated its properties, along with the associated current algebra, which constitutes a new  $\Z2$-graded  extension of the super-Virasoro algebra. 
This was done within the framework of Polyakov's soldering and the zero-curvature formulation. 
Explicit solutions of the derived equation were constructed by extending the method developed by Leznov and Saveliev. An auto-B\"acklund transformation was presented, along with a B\"acklund transformation to the free equation.

The graded extension of the super-Virasoro algebra was defined as a Poisson-Lie algebra, and we observed the presence of a non-vanishing [00]-graded central charge, despite the fact that the present theory is formulated within the framework of classical field theory. 
We pointed out that three inequivalent (anti)periodic boundary conditions are admissible in the mode expansion of the $\Z2$-graded currents. As a consequence, we obtained three distinct $\Z2$-graded extensions of the super-Virasoro algebra. However, the question of whether these algebras are equivalent remains an open problem.

The present work is based on a $\Z2$-graded extension of $\osp(1|2)$, and we have considered only [00] and [10]-graded superspace coordinates. This suggests the possibility of further $\Z2$-graded extensions of the super-Liouville equation. For instance, one may consider alternative $\Z2$-graded extensions of $ \osp(1|2)$, as the algebra admits some inequivalent $\Z2$-graded extensions \cite{rw2,GrJa,StoVDJ,StoVDJ3,StoVDJosp,RY}. It is also possible to formulate the theory on a superspace that includes [11]-graded coordinates. The square of the [11]-graded coordinate can be regarded as an emergent [00]-graded coordinate \cite{aizt,NARIint,NARITT2D}. Therefore, incorporating [11]-graded coordinates into the superspace may lead to integrable systems formulated in higher-dimensional spacetime.

Another interesting direction for future research is the study of integrable systems based on higher-rank $\Z2$-graded superalgebras such as $\Z2$-graded version of $\sl(2|1), \osp(2|2)$ and their affine extensions.  
Repeating the present analysis within a simplified superspace, as considered in this work, may lead to integrable systems that go beyond the $\Z2$-super-Liouville equation, since higher-rank algebras allow for the introduction of additional fields and interactions associated with the increasing number of simple roots. 
In this way, $\Z2$-graded superalgebras are expected to give rise to a rich landscape of novel integrable systems.

\section*{Acknowledgments}

The authors are grateful to Z. Kuznetsova for her contributions at the early stage of this work and for the stimulating and fruitful discussions throughout the project. 
This work was partially supported by the Global Strategy Fund of OMU. 
N. A. and R. I. are deeply grateful to CBPF and UFABC, where this work was completed, for hospitality. 
N. A. is supported by JSPS KAKENHI Grant Number JP23K03217. 
R. I. is supported by JST SPRING, Grant Number JPMJSP2139. 
F. T. is supported by
CNPq (PQ grant 308846/2021-4).


\end{document}